\documentclass[12pt]{article}
\usepackage[bottom=2cm, top=2cm, left=2cm, right=2cm]{geometry}
\usepackage{color}

\usepackage{multirow}
\usepackage{subfigure}
\usepackage{amsmath}
\usepackage[utf8]{inputenc}
\usepackage{standalone}
\usepackage{tikz}
\usepackage{amssymb}
\usepackage{eurosym}
\usepackage{hyperref}
\usepackage{caption}
\definecolor{b}{rgb}{0,0,.4}	
\definecolor{g}{rgb}{0,.3,0}	
\definecolor{n}{rgb}{0,0,0}	
\definecolor{h}{rgb}{0.4,0.2,0.2}	
\definecolor{v}{rgb}{0.2,0.6,0}

\DeclareMathOperator*{\argmin}{arg\,min}	

\newcommand{\ov}\overline

\newcommand{\wtilde}{\widetilde}

\definecolor{gray}{rgb}{0.5,0.5,0.5}
\definecolor{red}{rgb}{0.8,0,0}
\definecolor{dred}{rgb}{0.5,0,0}
\definecolor{blue}{rgb}{0,0.1,1}
\definecolor{dblue}{rgb}{0,0.1,0.6}
\definecolor{cyan}{rgb}{0,0.7,.2}
\definecolor{dcyan}{rgb}{0,0.5,.5}

\newcommand{\WSup}{\text{WSup}}
\newcommand{\WDem}{\text{WDem}}
\newcommand{\WDemInv}{\text{WDem}^{-1}}
\newcommand{\WSupInv}{\text{WSup}^{-1}}
\newcommand{\DemInv}{\text{Dem}^{-1}}
\newcommand{\SupInv}{\text{Sup}^{-1}}
\newcommand{\pmin}{p_{\min}}
\newcommand{\pmax}{p_{\max}}
\newcommand{\pupmax}{{(p^U, \pmax]}}
\newcommand{\pminpu}{{[\pmin, p^U)}}
\newcommand{\FSup}{\text{FSup}}
\newcommand{\FDem}{\text{FDem}}
\newcommand{\FDemInv}{\text{FDem}^{-1}}
\newcommand{\FSupInv}{\text{FSup}^{-1}}
\newcommand{\ione}{\mathbbm{1}}

\usetikzlibrary{arrows,positioning,shadows, calc, intersections, fit, automata}
\usetikzlibrary{arrows.meta}
\usetikzlibrary{decorations.pathreplacing}

 \usepackage{bbm}
\begin{document}
	\begin{center}
	{\large \textbf{Determining Fundamental Supply and Demand Curves in a Wholesale Electricity Market}}
	
	Sergei Kulakov\footnote{University of Duisburg-Essen, corresponding author. Email: sergei.kulakov@uni-due.de.} \& Florian Ziel\footnote{University of Duisburg-Essen.}

\end{center}
\begin{abstract}

 	In this paper we develop a novel method of wholesale electricity market modeling. Our optimization-based model decomposes wholesale supply and demand curves into buy and sell orders of individual market participants. In doing so, the model detects and removes arbitrage orders. As a result, we construct an innovative fundamental model of a wholesale electricity market. First, our fundamental demand curve has a unique composition. The demand curve lies {in between} the wholesale demand curve and a perfectly inelastic demand curve. Second, our fundamental supply and demand curves contain only actual (i.e. non-arbitrage) transactions with physical assets on buy and sell sides. Third, these transactions are designated to one of the three groups of wholesale electricity market participants: retailers, suppliers, or utility companies. To evaluate the performance of our model, we use the German wholesale market data. Our fundamental model yields a more precise approximation of the actual load values than a model with perfectly inelastic demand. Moreover, we conduct a study of wholesale demand elasticities. The obtained conclusions regarding wholesale demand elasticity are consistent with the existing academic literature. 
 	
 	
	
\end{abstract}

\section{Introduction}
The shift to renewable power is accelerating at a growing pace. Worldwide energy mix becomes increasingly reliant on wind and solar resources. Their apparent economic efficiency lures investments, while their environmental benefits attract vocal public support. Coping with their variability, however, is one of the major challenges of modern policymakers and electricity grid operators. 

Even before the onset of green technologies it was clear that successful integration of renewables requires a thorough revision and redesign of current energy infrastructure. As e.g. \cite{turner1999realizable} or \cite{boyle2004renewable} suggest, the issue of variability of green power can be tackled in multiple ways. Among those ways are, of course, a greater reliance on energy storage technologies (see e.g. \cite{carrasco2006power}, \cite{ibrahim2008energy}, or \cite{dunn2011electrical}), development of decentralized and smart grids (see e.g. \cite{heier2014grid}, \cite{liserre2010future}, \cite{mcdaniel2009security}l or \cite{kempton2005vehicle}) and, more importantly for the present paper,  a more advanced demand side management (see e.g. \cite{palensky2011demand}, \cite{mohsenian2010autonomous}, \cite{siano2014demand}).

{Modeling and forecasting in energy markets thus became an increasingly important and complex task (see e.g. \cite{weron2007modeling} or \cite{burger2008managing}). Models vary in their degree of sophistication, and focus on a whole multitude of aspects. In turn, in this paper we elaborate an optimization-based fundamental model of a wholesale electricity market. More specifically, our model is based on econometric manipulations with wholesale electricity supply and demand curves. Important is the fact that these manipulations are rooted into determining a unique composition of the fundamental demand curve. }

Our demand curve lies "in between" the wholesale demand curve (which incorporates many arbitrage orders) and a perfectly inelastic demand curve (which is an extreme assumption). In other words, our demand retains some elasticities, but eliminates arbitrage orders. {More specifically, our demand incorporates only actual transaction with physical assess on both buy and sell sides.} As a result, the demand curve in our fundamental model constitutes a better approximation of the true demand curve in a wholesale electricity market, i.e. of a curve which corresponds to the actual electricity load.


The paper is organized as follows. The remainder of the present section first provides a literature review. Then, subsections \ref{InelastSubs} and \ref{ASDWM} elaborate on details of the papers by \cite{coulon2014hourly} and \cite{knaut2016hourly}, respectively. These papers are worthy particular attention because they lay the foundation for our model. Finally, the introduction is concluded with a general overview of our contribution. Section \ref{ModelMain} is the description of our model. Subsection \ref{ModelIntuition} comments on general intuition behind our idea. Subsection \ref{ModelDescription} is devoted to the technical summary of the model. Subsection \ref{ModelToyTest} shows an application of our model to imaginary data. Institutional framework of the German electricity market as well as specifications of the data are outlined in section \ref{Data}. Section \ref{Results} discusses the obtained results. Section \ref{Conclusion} is the conclusion.

\subsection{Literature review}


{As was mentioned before, the kernel of our model is built on determining the composition of the fundamental demand curve. Therefore, to keep the literature review relatively concise, its focus will be placed on three main topics: approaches to modeling residential and industrial demand, discussion of literature on wholesale demand elasticities, and econometric modeling of wholesale supply and demand curves. Special attention will be paid to works by \cite{coulon2014hourly} and \cite{knaut2016hourly} because they lay the foundation for our research. }

Residential and industrial demand for energy has been analyzed extensively in academic literature. Therefore, multiple approaches to modeling demand in electricity markets exist. Following \cite{labandeira2012estimation}, many of these approaches payed particular attention to estimating demand elasticity. Not surprisingly so, as \cite{kirschen2003demand} suggest. Knowing price sensitivity of private households and industry players can be of great utility to energy producers and grid operators. Moreover, \cite{albadi2008summary} show that the success of different demand response programs and the corresponding cost-cuttings can be highly dependent on the accuracy of the demand elasticity estimation.

\cite{labandeira2012estimation} and \cite{bigerna2014electricity} argue that there are two main ways to model electricity demand. The former one relies on macroeconomic data to construct aggregate econometric models. The data may include electricity prices, income levels and climatic conditions. This approach was followed by e.g. \cite{narayan2005residential}, \cite{narayan2007electricity}, \cite{bernstein2006regional}, or \cite{holtedahl2004residential} to study residential demand, while e.g. \cite{kamerschen2004demand} and \cite{paul2009partial}, and \cite{taylor200524} analyzed industrial and aggregated demand besides residential one. 

The second way is to focus on microeconomic data or consumer surveys. A model may thus be based on characteristics, sizes, types, and preferences of households and companies. Among papers which restricted to this approach are e.g. \cite{alberini2011residential}, \cite{fuks2008applying}, \cite{leth2002micro}, \cite{labandeira2006residential}, \cite{fell2014new}, \cite{krishnamurthy2015cross} or \cite{schulte2017price} who studied residential elasticities. Industrial demand is investigated in, for example, \cite{woodland1993micro} and \cite{bardazzi2015manufacturing}. 

On the contrary, the body of academic literature on modeling wholesale demand is rather scarce. Moreover, papers in this field may follow slightly different modeling approaches. A more traditional method is used by e.g. \cite{lijesen2007real} who build up their model based on lagged day-ahead prices and temperature data. On the other hand, \cite{bonte2015price} use wind speed as a main proxy for demand elasticity. So do \cite{knaut2016hourly}. \cite{bigerna2014electricity} pursue yet another novel approach. They critique a common practice of determining residual demand and estimate demand elasticity from the available wholesale bid data. 

The present paper will be similar to the latter one in a sense that our modeling approach is also rooted into wholesale market data. However, we, too, develop a new and rather unconventional approach. We will focus solely on manipulations with wholesale supply and demand curves. 

Theoretical background of our paper thus stems from the field of econometric modeling of auction curves in a wholesale electricity market. \cite{barlow2002diffusion} were among the first researchers who attempted to construct a model based on real-world electricity auction data. \cite{buzoianu2012dynamic} study Californian electricity prices on the grounds of latent supply and demand curves. To estimate the curves, they exploit  temperature data, seasonality factors and gas availability. Furthermore, structural approaches have also be undertaken. \cite{carmona2013electricity} and \cite{howison2009stochastic} use bid stack to derive electricity spot prices on the grounds of power demand and prices of generating fuels. 

Wholesale auction curves were analyzed and manipulated in the following papers. \cite{ziel2016electricity} use day-ahead EEX auction curves to forecast German day-ahead electricity pries. Instead of analyzing the price time series, the authors of the paper suggest to predict auction curves in their entirety. The intersection of the predicted curves, of course, coincides with the price forecast. \cite{dillig2016impact} try to quantify the impact of renewable energies on electricity prices in the German market. Their model is also based on the EEX curves. To obtain the results, the authors of the paper add or subtract amounts of renewable power generation from the initial auction curves data. In other words, their model shifts one of the curves depending on the supply of renewable energy. A similar technique is employed in \cite{kulakov2019impact}. By shifting either the wholesale supply or demand curve, a non-linear impact of renewable energy generation on intra-day electricity prices is shown. The higher the renewable generation is, the stronger the prices in an intra-day market will react to the changes.

 Finally, the papers written by \cite{coulon2014hourly} and \cite{knaut2016hourly} require our particular attention because they build up the foundation for our research. These papers will thus be  discussed at length in the forthcoming two sections. 

\subsubsection{Transformation of the wholesale auction curves} \label{InelastSubs} 
\begin{figure}[h]
	\centering
	\vspace{-1cm}
	\scalebox{0.9}{\input{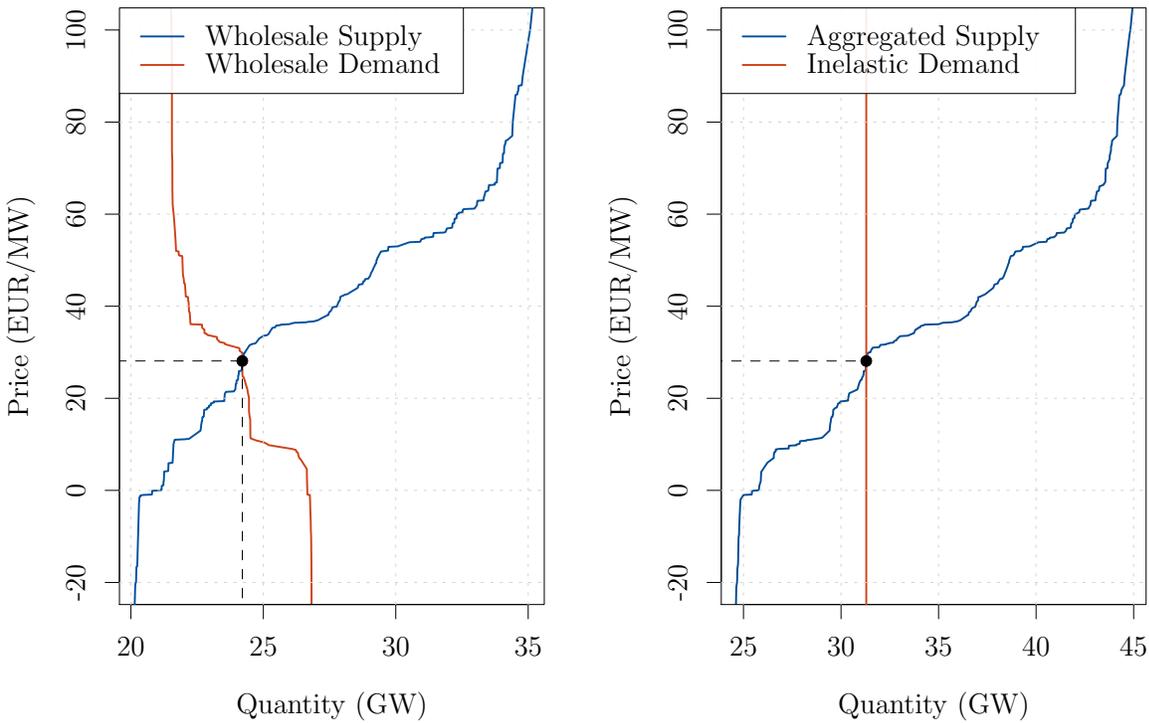}}		
	\caption{A wholesale market equilibrium in the EPEX SPOT SE on 2017-01-02 01:00:00 (left plot) vs. its manipulated form with an inelastic demand curve (right plot)}
	\label{FIG5}
\end{figure}

Of pivotal importance for the present study is a concept elaborated by \cite{coulon2014hourly}. The concept suggests that it is possible to transform wholesale supply and demand curves such that the demand curve becomes inelastic. Noteworthy is the fact that the equilibrium price remains the same before and after the transformation. On the contrary, the equilibrium volume increases. Figure \ref{FIG5} provides a graphical representation of the concept's functioning. The left hand side of the Figure shows the actual wholesale market auction curves observed in the German EPEX SPOT SE on 2017-01-01 at 00:00:00. The right hand side of the Figure depicts a transformed version of the two curves with perfectly inelastic demand. 

\cite{coulon2014hourly} elaborate on the intuition behind their idea at length. Following their reasoning, there are only few price-sensitive consumers in power markets. On the other hand, the demand curve in the wholesale market is elastic. At first, this contradiction seems counterintutive. However, besides the wholesale market, there exist another big bilateral venue where power suppliers are connected directly with their consumers. Therefore, whenever the prices in the wholesale market are lower than those in the bilateral one, multiple market participants {discover arbitrage opportunities and} decide to engage into speculation. They try to purchase electricity instead of producing it. Naturally, speculation leads to a surge in price sensitivity. 

Hence, having two parallel markets may lead to considerable complications for energy sector modeling. To avoid them, \cite{coulon2014hourly} suggest shifting the entire elasticity from the demand to the supply side. As a result, a perfectly inelastic demand curve will be obtained. Moreover, both the transformed supply and the demand curves are more stable and predictable because many orders in a wholesale market are of speculative nature. 

\subsubsection{Aggregated Supply and Demand case and the Wholesale Market} \label{ASDWM}
The kernel of our model stems from a market transformation carried out by \cite{knaut2016hourly}. However, intuition behind the transformation has not been elaborated in the original paper. Therefore, prior to explaining the model itself, we have to comment on functioning of its underpinnings. 

First of all, we introduce a toy example of a fictional electricity market. Describing the model on the grounds of the toy example appears to be more straightforward and much simpler. We assume an economy populated by three market participants: a Utility company, a Retailer, and a Supplier. The agents differ in their abilities to interact with the market. Utility can post both bid and ask orders, Retailer can only demand electricity from the market, and the choice of Supplier is restricted only to bid orders. Of course, the number of market participants can easily be expanded. In our case, however, each of them can be treated as the respective generalized representation of a country's power utility companies, electricity retailers, and electricity suppliers.

\begin{figure}
	\parbox{2.2in}{%
		\hspace{-0.5cm}\subfigure[\normalsize Utility  \label{UtilityFig}]{\scalebox{1}{
\begin{tikzpicture}[x=1pt,y=1pt]
\definecolor{fillColor}{RGB}{255,255,255}
\path[use as bounding box,fill=fillColor,fill opacity=0.00] (0,0) rectangle (144.54,216.81);
\begin{scope}
\path[clip] (  0.00,  0.00) rectangle (144.54,216.81);
\definecolor{drawColor}{RGB}{0,0,0}

\path[draw=drawColor,line width= 0.4pt,line join=round,line cap=round] ( 49.20, 72.18) -- ( 49.20,156.63);

\path[draw=drawColor,line width= 0.4pt,line join=round,line cap=round] ( 49.20, 72.18) -- ( 43.20, 72.18);

\path[draw=drawColor,line width= 0.4pt,line join=round,line cap=round] ( 49.20, 86.25) -- ( 43.20, 86.25);

\path[draw=drawColor,line width= 0.4pt,line join=round,line cap=round] ( 49.20,100.33) -- ( 43.20,100.33);

\path[draw=drawColor,line width= 0.4pt,line join=round,line cap=round] ( 49.20,114.41) -- ( 43.20,114.41);

\path[draw=drawColor,line width= 0.4pt,line join=round,line cap=round] ( 49.20,128.48) -- ( 43.20,128.48);

\path[draw=drawColor,line width= 0.4pt,line join=round,line cap=round] ( 49.20,142.56) -- ( 43.20,142.56);

\path[draw=drawColor,line width= 0.4pt,line join=round,line cap=round] ( 49.20,156.63) -- ( 43.20,156.63);

\node[text=drawColor,rotate= 90.00,anchor=base,inner sep=0pt, outer sep=0pt, scale=  1.00] at ( 34.80, 72.18) {-20};

\node[text=drawColor,rotate= 90.00,anchor=base,inner sep=0pt, outer sep=0pt, scale=  1.00] at ( 34.80,100.33) {20};

\node[text=drawColor,rotate= 90.00,anchor=base,inner sep=0pt, outer sep=0pt, scale=  1.00] at ( 34.80,128.48) {60};

\node[text=drawColor,rotate= 90.00,anchor=base,inner sep=0pt, outer sep=0pt, scale=  1.00] at ( 34.80,156.63) {100};

\path[draw=drawColor,line width= 0.4pt,line join=round,line cap=round] ( 49.20, 61.20) --
	(119.34, 61.20) --
	(119.34,167.61) --
	( 49.20,167.61) --
	( 49.20, 61.20);
\end{scope}
\begin{scope}
\path[clip] (  0.00,  0.00) rectangle (144.54,216.81);
\definecolor{drawColor}{RGB}{0,0,0}

\node[text=drawColor,anchor=base,inner sep=0pt, outer sep=0pt, scale=  0.92] at ( 84.27,189.04) {\bfseries Utility};

\node[text=drawColor,anchor=base,inner sep=0pt, outer sep=0pt, scale=  1.00] at ( 84.27, 15.60) {Quantity (MW)};

\node[text=drawColor,rotate= 90.00,anchor=base,inner sep=0pt, outer sep=0pt, scale=  1.00] at ( 10.80,114.41) {Price (EUR/MW)};
\end{scope}
\begin{scope}
\path[clip] (  0.00,  0.00) rectangle (144.54,216.81);
\definecolor{drawColor}{RGB}{0,0,0}

\path[draw=drawColor,line width= 0.4pt,line join=round,line cap=round] ( 51.80, 61.20) -- (119.34, 61.20);

\path[draw=drawColor,line width= 0.4pt,line join=round,line cap=round] ( 51.80, 61.20) -- ( 51.80, 55.20);

\path[draw=drawColor,line width= 0.4pt,line join=round,line cap=round] ( 70.35, 61.20) -- ( 70.35, 55.20);

\path[draw=drawColor,line width= 0.4pt,line join=round,line cap=round] ( 88.91, 61.20) -- ( 88.91, 55.20);

\path[draw=drawColor,line width= 0.4pt,line join=round,line cap=round] (107.46, 61.20) -- (107.46, 55.20);

\node[text=drawColor,anchor=base,inner sep=0pt, outer sep=0pt, scale=  1.00] at ( 51.80, 39.60) {0};

\node[text=drawColor,anchor=base,inner sep=0pt, outer sep=0pt, scale=  1.00] at ( 70.35, 39.60) {20};

\node[text=drawColor,anchor=base,inner sep=0pt, outer sep=0pt, scale=  1.00] at ( 88.91, 39.60) {40};

\node[text=drawColor,anchor=base,inner sep=0pt, outer sep=0pt, scale=  1.00] at (107.46, 39.60) {60};
\end{scope}
\begin{scope}
\path[clip] ( 49.20, 61.20) rectangle (119.34,167.61);
\definecolor{drawColor}{RGB}{0,107,69}

\path[draw=drawColor,line width= 1.2pt,line join=round,line cap=round] ( 98.19,121.44) --
	( 88.91,121.44);

\path[draw=drawColor,line width= 0.4pt,line join=round,line cap=round] ( 98.19,121.44) --
	( 98.19,216.81);

\path[draw=drawColor,line width= 1.2pt,line join=round,line cap=round] ( 88.91,107.37) --
	( 79.63,107.37);

\path[draw=drawColor,line width= 0.4pt,line join=round,line cap=round] ( 88.91,107.37) --
	( 88.91,121.44);

\path[draw=drawColor,line width= 1.2pt,line join=round,line cap=round] ( 79.63, 86.25) --
	( 70.35, 86.25);

\path[draw=drawColor,line width= 0.4pt,line join=round,line cap=round] ( 79.63, 86.25) --
	( 79.63,107.37);

\path[draw=drawColor,line width= 1.2pt,line join=round,line cap=round] ( 70.35, 79.22) --
	( 61.08, 79.22);

\path[draw=drawColor,line width= 0.4pt,line join=round,line cap=round] ( 70.35, 79.22) --
	( 70.35, 86.25);

\path[draw=drawColor,line width= 1.2pt,line join=round,line cap=round] ( 61.08, 72.18) --
	( 51.80, 72.18);

\path[draw=drawColor,line width= 0.4pt,line join=round,line cap=round] ( 61.08, 72.18) --
	( 61.08, 79.22);
\definecolor{drawColor}{RGB}{245,187,0}

\path[draw=drawColor,line width= 1.2pt,line join=round,line cap=round] ( 88.91, 72.18) --
	( 79.63, 72.18);

\path[draw=drawColor,line width= 0.4pt,line join=round,line cap=round] ( 88.91, 72.18) --
	( 88.91,  0.00);

\path[draw=drawColor,line width= 1.2pt,line join=round,line cap=round] ( 79.63,156.63) --
	( 51.80,156.63);

\path[draw=drawColor,line width= 0.4pt,line join=round,line cap=round] ( 79.63,156.63) --
	( 79.63, 72.18);
\definecolor{drawColor}{RGB}{211,211,211}

\path[draw=drawColor,line width= 0.4pt,dash pattern=on 1pt off 3pt ,line join=round,line cap=round] ( 51.80, 61.20) -- ( 51.80,167.61);

\path[draw=drawColor,line width= 0.4pt,dash pattern=on 1pt off 3pt ,line join=round,line cap=round] ( 61.08, 61.20) -- ( 61.08,167.61);

\path[draw=drawColor,line width= 0.4pt,dash pattern=on 1pt off 3pt ,line join=round,line cap=round] ( 70.35, 61.20) -- ( 70.35,167.61);

\path[draw=drawColor,line width= 0.4pt,dash pattern=on 1pt off 3pt ,line join=round,line cap=round] ( 79.63, 61.20) -- ( 79.63,167.61);

\path[draw=drawColor,line width= 0.4pt,dash pattern=on 1pt off 3pt ,line join=round,line cap=round] ( 88.91, 61.20) -- ( 88.91,167.61);

\path[draw=drawColor,line width= 0.4pt,dash pattern=on 1pt off 3pt ,line join=round,line cap=round] ( 98.19, 61.20) -- ( 98.19,167.61);

\path[draw=drawColor,line width= 0.4pt,dash pattern=on 1pt off 3pt ,line join=round,line cap=round] (107.46, 61.20) -- (107.46,167.61);

\path[draw=drawColor,line width= 0.4pt,dash pattern=on 1pt off 3pt ,line join=round,line cap=round] (116.74, 61.20) -- (116.74,167.61);

\path[draw=drawColor,line width= 0.4pt,dash pattern=on 1pt off 3pt ,line join=round,line cap=round] ( 49.20, 72.18) -- (119.34, 72.18);

\path[draw=drawColor,line width= 0.4pt,dash pattern=on 1pt off 3pt ,line join=round,line cap=round] ( 49.20, 86.25) -- (119.34, 86.25);

\path[draw=drawColor,line width= 0.4pt,dash pattern=on 1pt off 3pt ,line join=round,line cap=round] ( 49.20,100.33) -- (119.34,100.33);

\path[draw=drawColor,line width= 0.4pt,dash pattern=on 1pt off 3pt ,line join=round,line cap=round] ( 49.20,114.41) -- (119.34,114.41);

\path[draw=drawColor,line width= 0.4pt,dash pattern=on 1pt off 3pt ,line join=round,line cap=round] ( 49.20,128.48) -- (119.34,128.48);

\path[draw=drawColor,line width= 0.4pt,dash pattern=on 1pt off 3pt ,line join=round,line cap=round] ( 49.20,142.56) -- (119.34,142.56);

\path[draw=drawColor,line width= 0.4pt,dash pattern=on 1pt off 3pt ,line join=round,line cap=round] ( 49.20,156.63) -- (119.34,156.63);
\definecolor{drawColor}{RGB}{0,0,0}

\path[draw=drawColor,line width= 0.4pt,dash pattern=on 4pt off 4pt ,line join=round,line cap=round] ( 49.20, 86.25) -- (119.34, 86.25);
\end{scope}
\end{tikzpicture}}}}
	\begin{minipage}{1.6in}%
		\hspace{-1.3cm}\subfigure[\normalsize Retailer \label{RetailerFig}]{\scalebox{1}{
\begin{tikzpicture}[x=1pt,y=1pt]
\definecolor{fillColor}{RGB}{255,255,255}
\path[use as bounding box,fill=fillColor,fill opacity=0.00] (0,0) rectangle (144.54,216.81);
\begin{scope}
\path[clip] (  0.00,  0.00) rectangle (144.54,216.81);
\definecolor{drawColor}{RGB}{0,0,0}

\path[draw=drawColor,line width= 0.4pt,line join=round,line cap=round] ( 49.20, 72.18) -- ( 49.20,156.63);

\path[draw=drawColor,line width= 0.4pt,line join=round,line cap=round] ( 49.20, 72.18) -- ( 43.20, 72.18);

\path[draw=drawColor,line width= 0.4pt,line join=round,line cap=round] ( 49.20, 86.25) -- ( 43.20, 86.25);

\path[draw=drawColor,line width= 0.4pt,line join=round,line cap=round] ( 49.20,100.33) -- ( 43.20,100.33);

\path[draw=drawColor,line width= 0.4pt,line join=round,line cap=round] ( 49.20,114.41) -- ( 43.20,114.41);

\path[draw=drawColor,line width= 0.4pt,line join=round,line cap=round] ( 49.20,128.48) -- ( 43.20,128.48);

\path[draw=drawColor,line width= 0.4pt,line join=round,line cap=round] ( 49.20,142.56) -- ( 43.20,142.56);

\path[draw=drawColor,line width= 0.4pt,line join=round,line cap=round] ( 49.20,156.63) -- ( 43.20,156.63);

\node[text=drawColor,rotate= 90.00,anchor=base,inner sep=0pt, outer sep=0pt, scale=  1.00] at ( 34.80, 72.18) {-20};

\node[text=drawColor,rotate= 90.00,anchor=base,inner sep=0pt, outer sep=0pt, scale=  1.00] at ( 34.80,100.33) {20};

\node[text=drawColor,rotate= 90.00,anchor=base,inner sep=0pt, outer sep=0pt, scale=  1.00] at ( 34.80,128.48) {60};

\node[text=drawColor,rotate= 90.00,anchor=base,inner sep=0pt, outer sep=0pt, scale=  1.00] at ( 34.80,156.63) {100};

\path[draw=drawColor,line width= 0.4pt,line join=round,line cap=round] ( 49.20, 61.20) --
	(119.34, 61.20) --
	(119.34,167.61) --
	( 49.20,167.61) --
	( 49.20, 61.20);
\end{scope}
\begin{scope}
\path[clip] (  0.00,  0.00) rectangle (144.54,216.81);
\definecolor{drawColor}{RGB}{0,0,0}

\node[text=drawColor,anchor=base,inner sep=0pt, outer sep=0pt, scale=  0.92] at ( 84.27,189.04) {\bfseries Retailer};

\node[text=drawColor,anchor=base,inner sep=0pt, outer sep=0pt, scale=  1.00] at ( 84.27, 15.60) {Quantity (MW)};

\node[text=drawColor,rotate= 90.00,anchor=base,inner sep=0pt, outer sep=0pt, scale=  1.00] at ( 10.80,114.41) {Price (EUR/MW)};
\end{scope}
\begin{scope}
\path[clip] (  0.00,  0.00) rectangle (144.54,216.81);
\definecolor{drawColor}{RGB}{0,0,0}

\path[draw=drawColor,line width= 0.4pt,line join=round,line cap=round] ( 51.80, 61.20) -- (119.34, 61.20);

\path[draw=drawColor,line width= 0.4pt,line join=round,line cap=round] ( 51.80, 61.20) -- ( 51.80, 55.20);

\path[draw=drawColor,line width= 0.4pt,line join=round,line cap=round] ( 70.35, 61.20) -- ( 70.35, 55.20);

\path[draw=drawColor,line width= 0.4pt,line join=round,line cap=round] ( 88.91, 61.20) -- ( 88.91, 55.20);

\path[draw=drawColor,line width= 0.4pt,line join=round,line cap=round] (107.46, 61.20) -- (107.46, 55.20);

\node[text=drawColor,anchor=base,inner sep=0pt, outer sep=0pt, scale=  1.00] at ( 51.80, 39.60) {0};

\node[text=drawColor,anchor=base,inner sep=0pt, outer sep=0pt, scale=  1.00] at ( 70.35, 39.60) {20};

\node[text=drawColor,anchor=base,inner sep=0pt, outer sep=0pt, scale=  1.00] at ( 88.91, 39.60) {40};

\node[text=drawColor,anchor=base,inner sep=0pt, outer sep=0pt, scale=  1.00] at (107.46, 39.60) {60};
\end{scope}
\begin{scope}
\path[clip] ( 49.20, 61.20) rectangle (119.34,167.61);
\definecolor{drawColor}{RGB}{139,48,13}

\path[draw=drawColor,line width= 1.2pt,line join=round,line cap=round] (107.46, 72.18) --
	( 98.19, 72.18);

\path[draw=drawColor,line width= 0.4pt,line join=round,line cap=round] (107.46, 72.18) --
	(107.46,  0.00);

\path[draw=drawColor,line width= 1.2pt,line join=round,line cap=round] ( 98.19, 79.22) --
	( 88.91, 79.22);

\path[draw=drawColor,line width= 0.4pt,line join=round,line cap=round] ( 98.19, 79.22) --
	( 98.19, 72.18);

\path[draw=drawColor,line width= 1.2pt,line join=round,line cap=round] ( 88.91,114.41) --
	( 70.35,114.41);

\path[draw=drawColor,line width= 0.4pt,line join=round,line cap=round] ( 88.91,114.41) --
	( 88.91, 79.22);

\path[draw=drawColor,line width= 1.2pt,line join=round,line cap=round] ( 70.35,149.59) --
	( 61.08,149.59);

\path[draw=drawColor,line width= 0.4pt,line join=round,line cap=round] ( 70.35,149.59) --
	( 70.35,114.41);

\path[draw=drawColor,line width= 1.2pt,line join=round,line cap=round] ( 61.08,156.63) --
	( 51.80,156.63);

\path[draw=drawColor,line width= 0.4pt,line join=round,line cap=round] ( 61.08,156.63) --
	( 61.08,149.59);
\definecolor{drawColor}{RGB}{211,211,211}

\path[draw=drawColor,line width= 0.4pt,dash pattern=on 1pt off 3pt ,line join=round,line cap=round] ( 51.80, 61.20) -- ( 51.80,167.61);

\path[draw=drawColor,line width= 0.4pt,dash pattern=on 1pt off 3pt ,line join=round,line cap=round] ( 61.08, 61.20) -- ( 61.08,167.61);

\path[draw=drawColor,line width= 0.4pt,dash pattern=on 1pt off 3pt ,line join=round,line cap=round] ( 70.35, 61.20) -- ( 70.35,167.61);

\path[draw=drawColor,line width= 0.4pt,dash pattern=on 1pt off 3pt ,line join=round,line cap=round] ( 79.63, 61.20) -- ( 79.63,167.61);

\path[draw=drawColor,line width= 0.4pt,dash pattern=on 1pt off 3pt ,line join=round,line cap=round] ( 88.91, 61.20) -- ( 88.91,167.61);

\path[draw=drawColor,line width= 0.4pt,dash pattern=on 1pt off 3pt ,line join=round,line cap=round] ( 98.19, 61.20) -- ( 98.19,167.61);

\path[draw=drawColor,line width= 0.4pt,dash pattern=on 1pt off 3pt ,line join=round,line cap=round] (107.46, 61.20) -- (107.46,167.61);

\path[draw=drawColor,line width= 0.4pt,dash pattern=on 1pt off 3pt ,line join=round,line cap=round] (116.74, 61.20) -- (116.74,167.61);

\path[draw=drawColor,line width= 0.4pt,dash pattern=on 1pt off 3pt ,line join=round,line cap=round] ( 49.20, 72.18) -- (119.34, 72.18);

\path[draw=drawColor,line width= 0.4pt,dash pattern=on 1pt off 3pt ,line join=round,line cap=round] ( 49.20, 86.25) -- (119.34, 86.25);

\path[draw=drawColor,line width= 0.4pt,dash pattern=on 1pt off 3pt ,line join=round,line cap=round] ( 49.20,100.33) -- (119.34,100.33);

\path[draw=drawColor,line width= 0.4pt,dash pattern=on 1pt off 3pt ,line join=round,line cap=round] ( 49.20,114.41) -- (119.34,114.41);

\path[draw=drawColor,line width= 0.4pt,dash pattern=on 1pt off 3pt ,line join=round,line cap=round] ( 49.20,128.48) -- (119.34,128.48);

\path[draw=drawColor,line width= 0.4pt,dash pattern=on 1pt off 3pt ,line join=round,line cap=round] ( 49.20,142.56) -- (119.34,142.56);

\path[draw=drawColor,line width= 0.4pt,dash pattern=on 1pt off 3pt ,line join=round,line cap=round] ( 49.20,156.63) -- (119.34,156.63);
\end{scope}
\end{tikzpicture}}}
	\end{minipage}%
	\begin{minipage}{1.5in}%
		\hspace*{-0.4cm}\subfigure[\normalsize Supplier \label{SupplierFig}]{\scalebox{1}{
\begin{tikzpicture}[x=1pt,y=1pt]
\definecolor{fillColor}{RGB}{255,255,255}
\path[use as bounding box,fill=fillColor,fill opacity=0.00] (0,0) rectangle (144.54,216.81);
\begin{scope}
\path[clip] (  0.00,  0.00) rectangle (144.54,216.81);
\definecolor{drawColor}{RGB}{0,0,0}

\path[draw=drawColor,line width= 0.4pt,line join=round,line cap=round] ( 49.20, 72.18) -- ( 49.20,156.63);

\path[draw=drawColor,line width= 0.4pt,line join=round,line cap=round] ( 49.20, 72.18) -- ( 43.20, 72.18);

\path[draw=drawColor,line width= 0.4pt,line join=round,line cap=round] ( 49.20, 86.25) -- ( 43.20, 86.25);

\path[draw=drawColor,line width= 0.4pt,line join=round,line cap=round] ( 49.20,100.33) -- ( 43.20,100.33);

\path[draw=drawColor,line width= 0.4pt,line join=round,line cap=round] ( 49.20,114.41) -- ( 43.20,114.41);

\path[draw=drawColor,line width= 0.4pt,line join=round,line cap=round] ( 49.20,128.48) -- ( 43.20,128.48);

\path[draw=drawColor,line width= 0.4pt,line join=round,line cap=round] ( 49.20,142.56) -- ( 43.20,142.56);

\path[draw=drawColor,line width= 0.4pt,line join=round,line cap=round] ( 49.20,156.63) -- ( 43.20,156.63);

\node[text=drawColor,rotate= 90.00,anchor=base,inner sep=0pt, outer sep=0pt, scale=  1.00] at ( 34.80, 72.18) {-20};

\node[text=drawColor,rotate= 90.00,anchor=base,inner sep=0pt, outer sep=0pt, scale=  1.00] at ( 34.80,100.33) {20};

\node[text=drawColor,rotate= 90.00,anchor=base,inner sep=0pt, outer sep=0pt, scale=  1.00] at ( 34.80,128.48) {60};

\node[text=drawColor,rotate= 90.00,anchor=base,inner sep=0pt, outer sep=0pt, scale=  1.00] at ( 34.80,156.63) {100};

\path[draw=drawColor,line width= 0.4pt,line join=round,line cap=round] ( 49.20, 61.20) --
	(119.34, 61.20) --
	(119.34,167.61) --
	( 49.20,167.61) --
	( 49.20, 61.20);
\end{scope}
\begin{scope}
\path[clip] (  0.00,  0.00) rectangle (144.54,216.81);
\definecolor{drawColor}{RGB}{0,0,0}

\node[text=drawColor,anchor=base,inner sep=0pt, outer sep=0pt, scale=  0.92] at ( 84.27,189.04) {\bfseries Supplier};

\node[text=drawColor,anchor=base,inner sep=0pt, outer sep=0pt, scale=  1.00] at ( 84.27, 15.60) {Quantity (MW)};

\node[text=drawColor,rotate= 90.00,anchor=base,inner sep=0pt, outer sep=0pt, scale=  1.00] at ( 10.80,114.41) {Price (EUR/MW)};
\end{scope}
\begin{scope}
\path[clip] (  0.00,  0.00) rectangle (144.54,216.81);
\definecolor{drawColor}{RGB}{0,0,0}

\path[draw=drawColor,line width= 0.4pt,line join=round,line cap=round] ( 51.80, 61.20) -- (119.34, 61.20);

\path[draw=drawColor,line width= 0.4pt,line join=round,line cap=round] ( 51.80, 61.20) -- ( 51.80, 55.20);

\path[draw=drawColor,line width= 0.4pt,line join=round,line cap=round] ( 70.35, 61.20) -- ( 70.35, 55.20);

\path[draw=drawColor,line width= 0.4pt,line join=round,line cap=round] ( 88.91, 61.20) -- ( 88.91, 55.20);

\path[draw=drawColor,line width= 0.4pt,line join=round,line cap=round] (107.46, 61.20) -- (107.46, 55.20);

\node[text=drawColor,anchor=base,inner sep=0pt, outer sep=0pt, scale=  1.00] at ( 51.80, 39.60) {0};

\node[text=drawColor,anchor=base,inner sep=0pt, outer sep=0pt, scale=  1.00] at ( 70.35, 39.60) {20};

\node[text=drawColor,anchor=base,inner sep=0pt, outer sep=0pt, scale=  1.00] at ( 88.91, 39.60) {40};

\node[text=drawColor,anchor=base,inner sep=0pt, outer sep=0pt, scale=  1.00] at (107.46, 39.60) {60};
\end{scope}
\begin{scope}
\path[clip] ( 49.20, 61.20) rectangle (119.34,167.61);
\definecolor{drawColor}{RGB}{62,53,213}

\path[draw=drawColor,line width= 1.2pt,line join=round,line cap=round] (107.46,114.41) --
	( 98.19,114.41);

\path[draw=drawColor,line width= 0.4pt,line join=round,line cap=round] (107.46,114.41) --
	(107.46,216.81);

\path[draw=drawColor,line width= 1.2pt,line join=round,line cap=round] ( 98.19,100.33) --
	( 79.63,100.33);

\path[draw=drawColor,line width= 0.4pt,line join=round,line cap=round] ( 98.19,100.33) --
	( 98.19,114.41);

\path[draw=drawColor,line width= 1.2pt,line join=round,line cap=round] ( 79.63, 93.29) --
	( 70.35, 93.29);

\path[draw=drawColor,line width= 0.4pt,line join=round,line cap=round] ( 79.63, 93.29) --
	( 79.63,100.33);

\path[draw=drawColor,line width= 1.2pt,line join=round,line cap=round] ( 70.35, 86.25) --
	( 61.08, 86.25);

\path[draw=drawColor,line width= 0.4pt,line join=round,line cap=round] ( 70.35, 86.25) --
	( 70.35, 93.29);

\path[draw=drawColor,line width= 1.2pt,line join=round,line cap=round] ( 61.08, 79.22) --
	( 51.80, 79.22);

\path[draw=drawColor,line width= 0.4pt,line join=round,line cap=round] ( 61.08, 79.22) --
	( 61.08, 86.25);
\definecolor{drawColor}{RGB}{211,211,211}

\path[draw=drawColor,line width= 0.4pt,dash pattern=on 1pt off 3pt ,line join=round,line cap=round] ( 51.80, 61.20) -- ( 51.80,167.61);

\path[draw=drawColor,line width= 0.4pt,dash pattern=on 1pt off 3pt ,line join=round,line cap=round] ( 61.08, 61.20) -- ( 61.08,167.61);

\path[draw=drawColor,line width= 0.4pt,dash pattern=on 1pt off 3pt ,line join=round,line cap=round] ( 70.35, 61.20) -- ( 70.35,167.61);

\path[draw=drawColor,line width= 0.4pt,dash pattern=on 1pt off 3pt ,line join=round,line cap=round] ( 79.63, 61.20) -- ( 79.63,167.61);

\path[draw=drawColor,line width= 0.4pt,dash pattern=on 1pt off 3pt ,line join=round,line cap=round] ( 88.91, 61.20) -- ( 88.91,167.61);

\path[draw=drawColor,line width= 0.4pt,dash pattern=on 1pt off 3pt ,line join=round,line cap=round] ( 98.19, 61.20) -- ( 98.19,167.61);

\path[draw=drawColor,line width= 0.4pt,dash pattern=on 1pt off 3pt ,line join=round,line cap=round] (107.46, 61.20) -- (107.46,167.61);

\path[draw=drawColor,line width= 0.4pt,dash pattern=on 1pt off 3pt ,line join=round,line cap=round] (116.74, 61.20) -- (116.74,167.61);

\path[draw=drawColor,line width= 0.4pt,dash pattern=on 1pt off 3pt ,line join=round,line cap=round] ( 49.20, 72.18) -- (119.34, 72.18);

\path[draw=drawColor,line width= 0.4pt,dash pattern=on 1pt off 3pt ,line join=round,line cap=round] ( 49.20, 86.25) -- (119.34, 86.25);

\path[draw=drawColor,line width= 0.4pt,dash pattern=on 1pt off 3pt ,line join=round,line cap=round] ( 49.20,100.33) -- (119.34,100.33);

\path[draw=drawColor,line width= 0.4pt,dash pattern=on 1pt off 3pt ,line join=round,line cap=round] ( 49.20,114.41) -- (119.34,114.41);

\path[draw=drawColor,line width= 0.4pt,dash pattern=on 1pt off 3pt ,line join=round,line cap=round] ( 49.20,128.48) -- (119.34,128.48);

\path[draw=drawColor,line width= 0.4pt,dash pattern=on 1pt off 3pt ,line join=round,line cap=round] ( 49.20,142.56) -- (119.34,142.56);

\path[draw=drawColor,line width= 0.4pt,dash pattern=on 1pt off 3pt ,line join=round,line cap=round] ( 49.20,156.63) -- (119.34,156.63);
\end{scope}
\end{tikzpicture}}}
	\end{minipage}%
\centering
	
	\parbox{2.5in}{%
		\hspace{-0.5cm}\subfigure[\normalsize Aggregated supply and demand  \label{ASD}]{\scalebox{1}{\input{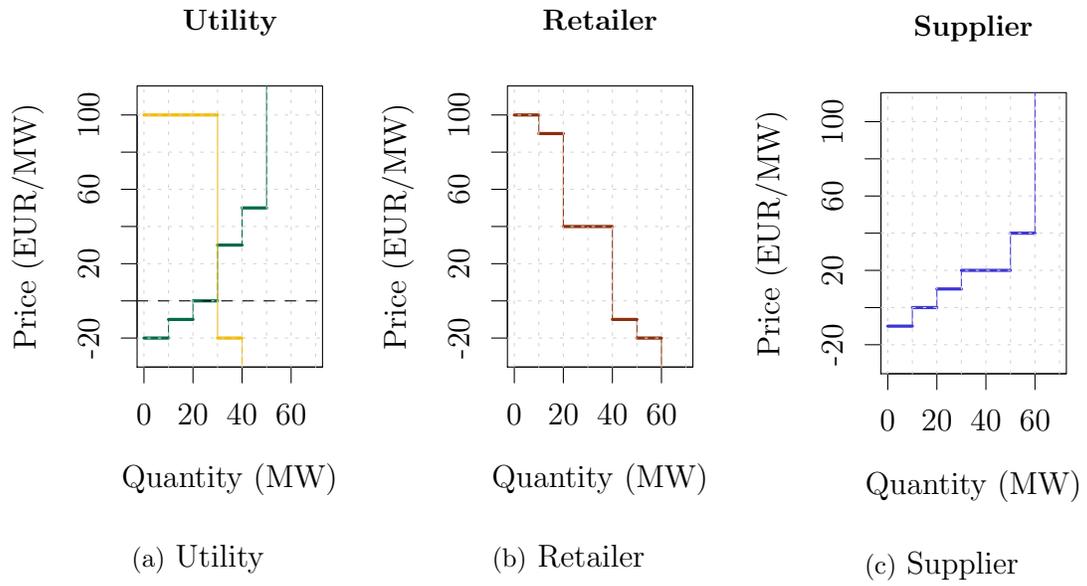}}}}
	\qquad
	\begin{minipage}{1.9in}%
		\hspace{-0.5cm}\hspace{-1.5cm}\subfigure[\normalsize Wholesale market \label{WM}]{\scalebox{1}{\input{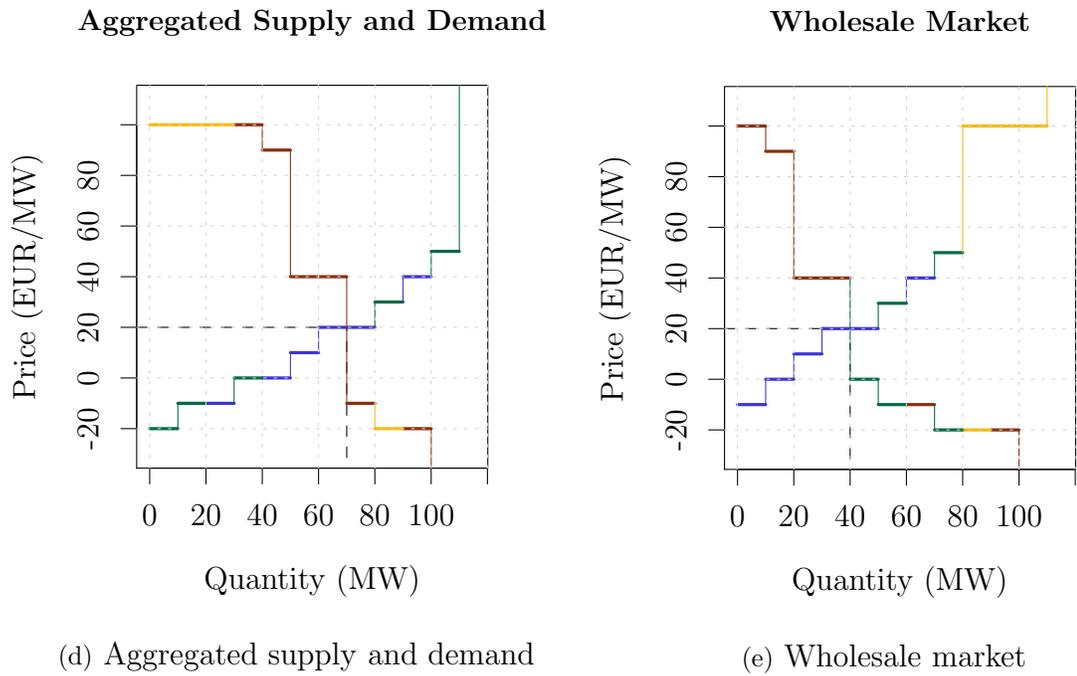}}}
	\end{minipage}%
	\vspace{0.2cm}
	\caption{A toy example of an electricity market}
	\label{FIGMARK}
\end{figure}

Individual supply and demand schedules of the market agents in our toy example are represented in Figures \ref{UtilityFig}, \ref{RetailerFig}, and \ref{SupplierFig}, respectively. Please note that the internal equilibrium price of Utility equals to zero. This price was selected because it allows us to demonstrate all features of our model more explicitly. Moreover, to ease further notation, each single step in the supply curves of the market participants will be referred to as a sell order. In turn, by a buy order we will denote a single step in the demand curves. 

As \cite{knaut2016hourly} suggest, traders' orders can be aggregated by means of two different techniques. The two resulting equilibria will be referred to as an Aggregated Supply and Demand case (ASD) and a Wholesale Market (WM). Even though the choice of a technique does non influence equilibrium prices, it does affect equilibrium volumes and the compositions of the market supply and demand curves. The focus of the present section will thus be placed on the discussion of these two techniques. 

Graphical depictions of both final equilibria are provided in Figures \ref{ASD} and \ref{WM}, respectively. Explaining how both solutions were obtained is another challenge we face at this stage. To simplify the forthcoming description, we create two auxiliary variables: a supply pool and a demand pool. Following their names, the variables will be used to collect buy and sell orders of the market participants.  Separation into two pools will become especially useful for the discussion of the more cumbersome WM case. 

Let us first concentrate on the ASD equilibrium and Figure \ref{ASD}. The demand curve in this case was built up as follows. First, we collected Utility's and Retailer's buy orders in the demand pool. Second, we sorted the orders in the demand pool according to their prices. Third, we plotted the obtained data. The supply curve in this case was then constructed analogously. First, we collected Utility's and Supplier's sell orders in the supply pool. Second, we sorted the orders in the supply pool according to their prices. Third, we plotted the obtained data.  The upper part Table \ref{TABLESUPPL} summarizes components of both the supply and demand pools in the ASD case.

\begin{figure}[h]
		\hspace*{1cm}
	\parbox{2.2in}{%
		\hspace{-0.5cm}\subfigure[\normalsize Utility Sectors \label{UtilitySectors}]{\scalebox{0.9}{\input{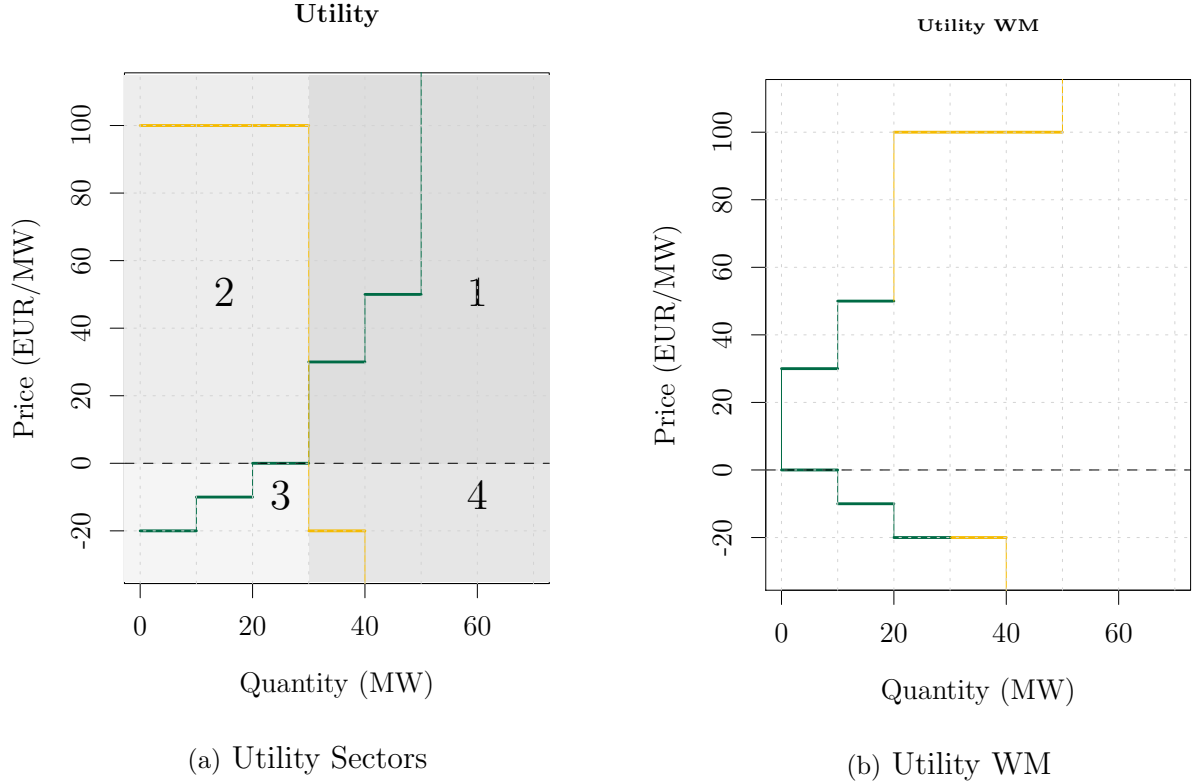}}}}
	\begin{minipage}{01in}%
		\hspace*{2.3cm}\subfigure[\normalsize Utility WM \label{UtilityWM}]{\scalebox{0.9}{
\begin{tikzpicture}[x=1pt,y=1pt]
\definecolor{fillColor}{RGB}{255,255,255}
\path[use as bounding box,fill=fillColor,fill opacity=0.00] (0,0) rectangle (252.94,325.21);
\begin{scope}
\path[clip] (  0.00,  0.00) rectangle (252.94,325.21);
\definecolor{drawColor}{RGB}{0,0,0}

\path[draw=drawColor,line width= 0.4pt,line join=round,line cap=round] ( 49.20, 83.36) -- ( 49.20,253.85);

\path[draw=drawColor,line width= 0.4pt,line join=round,line cap=round] ( 49.20, 83.36) -- ( 43.20, 83.36);

\path[draw=drawColor,line width= 0.4pt,line join=round,line cap=round] ( 49.20,111.78) -- ( 43.20,111.78);

\path[draw=drawColor,line width= 0.4pt,line join=round,line cap=round] ( 49.20,140.19) -- ( 43.20,140.19);

\path[draw=drawColor,line width= 0.4pt,line join=round,line cap=round] ( 49.20,168.61) -- ( 43.20,168.61);

\path[draw=drawColor,line width= 0.4pt,line join=round,line cap=round] ( 49.20,197.02) -- ( 43.20,197.02);

\path[draw=drawColor,line width= 0.4pt,line join=round,line cap=round] ( 49.20,225.44) -- ( 43.20,225.44);

\path[draw=drawColor,line width= 0.4pt,line join=round,line cap=round] ( 49.20,253.85) -- ( 43.20,253.85);

\node[text=drawColor,rotate= 90.00,anchor=base,inner sep=0pt, outer sep=0pt, scale=  1.00] at ( 34.80, 83.36) {-20};

\node[text=drawColor,rotate= 90.00,anchor=base,inner sep=0pt, outer sep=0pt, scale=  1.00] at ( 34.80,111.78) {0};

\node[text=drawColor,rotate= 90.00,anchor=base,inner sep=0pt, outer sep=0pt, scale=  1.00] at ( 34.80,140.19) {20};

\node[text=drawColor,rotate= 90.00,anchor=base,inner sep=0pt, outer sep=0pt, scale=  1.00] at ( 34.80,168.61) {40};

\node[text=drawColor,rotate= 90.00,anchor=base,inner sep=0pt, outer sep=0pt, scale=  1.00] at ( 34.80,197.02) {60};

\node[text=drawColor,rotate= 90.00,anchor=base,inner sep=0pt, outer sep=0pt, scale=  1.00] at ( 34.80,225.44) {80};

\node[text=drawColor,rotate= 90.00,anchor=base,inner sep=0pt, outer sep=0pt, scale=  1.00] at ( 34.80,253.85) {100};

\path[draw=drawColor,line width= 0.4pt,line join=round,line cap=round] ( 49.20, 61.20) --
	(227.75, 61.20) --
	(227.75,276.01) --
	( 49.20,276.01) --
	( 49.20, 61.20);
\end{scope}
\begin{scope}
\path[clip] (  0.00,  0.00) rectangle (252.94,325.21);
\definecolor{drawColor}{RGB}{0,0,0}

\node[text=drawColor,anchor=base,inner sep=0pt, outer sep=0pt, scale=  1.20] at (138.47,296.47) {\bfseries \tiny{Utility WM}};

\node[text=drawColor,anchor=base,inner sep=0pt, outer sep=0pt, scale=  1.00] at (138.47, 15.60) {Quantity (MW)};

\node[text=drawColor,rotate= 90.00,anchor=base,inner sep=0pt, outer sep=0pt, scale=  1.00] at ( 10.80,168.61) {Price (EUR/MW)};
\end{scope}
\begin{scope}
\path[clip] (  0.00,  0.00) rectangle (252.94,325.21);
\definecolor{drawColor}{RGB}{0,0,0}

\path[draw=drawColor,line width= 0.4pt,line join=round,line cap=round] ( 55.81, 61.20) -- (227.75, 61.20);

\path[draw=drawColor,line width= 0.4pt,line join=round,line cap=round] ( 55.81, 61.20) -- ( 55.81, 55.20);

\path[draw=drawColor,line width= 0.4pt,line join=round,line cap=round] (103.05, 61.20) -- (103.05, 55.20);

\path[draw=drawColor,line width= 0.4pt,line join=round,line cap=round] (150.28, 61.20) -- (150.28, 55.20);

\path[draw=drawColor,line width= 0.4pt,line join=round,line cap=round] (197.52, 61.20) -- (197.52, 55.20);

\node[text=drawColor,anchor=base,inner sep=0pt, outer sep=0pt, scale=  1.00] at ( 55.81, 39.60) {0};

\node[text=drawColor,anchor=base,inner sep=0pt, outer sep=0pt, scale=  1.00] at (103.05, 39.60) {20};

\node[text=drawColor,anchor=base,inner sep=0pt, outer sep=0pt, scale=  1.00] at (150.28, 39.60) {40};

\node[text=drawColor,anchor=base,inner sep=0pt, outer sep=0pt, scale=  1.00] at (197.52, 39.60) {60};
\end{scope}
\begin{scope}
\path[clip] ( 49.20, 61.20) rectangle (227.75,276.01);
\definecolor{drawColor}{RGB}{245,187,0}

\path[draw=drawColor,line width= 1.2pt,line join=round,line cap=round] (173.90,253.85) --
	(103.05,253.85);

\path[draw=drawColor,line width= 0.4pt,line join=round,line cap=round] (173.90,253.85) --
	(173.90,325.21);
\definecolor{drawColor}{RGB}{0,107,69}

\path[draw=drawColor,line width= 1.2pt,line join=round,line cap=round] (103.05,182.81) --
	( 79.43,182.81);
\definecolor{drawColor}{RGB}{245,187,0}

\path[draw=drawColor,line width= 0.4pt,line join=round,line cap=round] (103.05,182.81) --
	(103.05,253.85);
\definecolor{drawColor}{RGB}{0,107,69}

\path[draw=drawColor,line width= 1.2pt,line join=round,line cap=round] ( 79.43,154.40) --
	( 55.81,154.40);

\path[draw=drawColor,line width= 0.4pt,line join=round,line cap=round] ( 79.43,154.40) --
	( 79.43,182.81);
\definecolor{drawColor}{RGB}{245,187,0}

\path[draw=drawColor,line width= 1.2pt,line join=round,line cap=round] (150.28, 83.36) --
	(126.66, 83.36);

\path[draw=drawColor,line width= 0.4pt,line join=round,line cap=round] (150.28, 83.36) --
	(150.28,  0.00);
\definecolor{drawColor}{RGB}{0,107,69}

\path[draw=drawColor,line width= 1.2pt,line join=round,line cap=round] (126.66, 83.36) --
	(103.05, 83.36);
\definecolor{drawColor}{RGB}{245,187,0}

\path[draw=drawColor,line width= 0.4pt,line join=round,line cap=round] (126.66, 83.36) --
	(126.66, 83.36);
\definecolor{drawColor}{RGB}{0,107,69}

\path[draw=drawColor,line width= 1.2pt,line join=round,line cap=round] (103.05, 97.57) --
	( 79.43, 97.57);

\path[draw=drawColor,line width= 0.4pt,line join=round,line cap=round] (103.05, 97.57) --
	(103.05, 83.36);

\path[draw=drawColor,line width= 1.2pt,line join=round,line cap=round] ( 79.43,111.78) --
	( 55.81,111.78);

\path[draw=drawColor,line width= 0.4pt,line join=round,line cap=round] ( 79.43,111.78) --
	( 79.43, 97.57);
\definecolor{drawColor}{RGB}{211,211,211}

\path[draw=drawColor,line width= 0.4pt,dash pattern=on 1pt off 3pt ,line join=round,line cap=round] ( 55.81, 61.20) -- ( 55.81,276.01);

\path[draw=drawColor,line width= 0.4pt,dash pattern=on 1pt off 3pt ,line join=round,line cap=round] ( 79.43, 61.20) -- ( 79.43,276.01);

\path[draw=drawColor,line width= 0.4pt,dash pattern=on 1pt off 3pt ,line join=round,line cap=round] (103.05, 61.20) -- (103.05,276.01);

\path[draw=drawColor,line width= 0.4pt,dash pattern=on 1pt off 3pt ,line join=round,line cap=round] (126.66, 61.20) -- (126.66,276.01);

\path[draw=drawColor,line width= 0.4pt,dash pattern=on 1pt off 3pt ,line join=round,line cap=round] (150.28, 61.20) -- (150.28,276.01);

\path[draw=drawColor,line width= 0.4pt,dash pattern=on 1pt off 3pt ,line join=round,line cap=round] (173.90, 61.20) -- (173.90,276.01);

\path[draw=drawColor,line width= 0.4pt,dash pattern=on 1pt off 3pt ,line join=round,line cap=round] (197.52, 61.20) -- (197.52,276.01);

\path[draw=drawColor,line width= 0.4pt,dash pattern=on 1pt off 3pt ,line join=round,line cap=round] (221.13, 61.20) -- (221.13,276.01);

\path[draw=drawColor,line width= 0.4pt,dash pattern=on 1pt off 3pt ,line join=round,line cap=round] ( 49.20, 83.36) -- (227.75, 83.36);

\path[draw=drawColor,line width= 0.4pt,dash pattern=on 1pt off 3pt ,line join=round,line cap=round] ( 49.20,111.78) -- (227.75,111.78);

\path[draw=drawColor,line width= 0.4pt,dash pattern=on 1pt off 3pt ,line join=round,line cap=round] ( 49.20,140.19) -- (227.75,140.19);

\path[draw=drawColor,line width= 0.4pt,dash pattern=on 1pt off 3pt ,line join=round,line cap=round] ( 49.20,168.61) -- (227.75,168.61);

\path[draw=drawColor,line width= 0.4pt,dash pattern=on 1pt off 3pt ,line join=round,line cap=round] ( 49.20,197.02) -- (227.75,197.02);

\path[draw=drawColor,line width= 0.4pt,dash pattern=on 1pt off 3pt ,line join=round,line cap=round] ( 49.20,225.44) -- (227.75,225.44);

\path[draw=drawColor,line width= 0.4pt,dash pattern=on 1pt off 3pt ,line join=round,line cap=round] ( 49.20,253.85) -- (227.75,253.85);
\definecolor{drawColor}{RGB}{0,0,0}

\path[draw=drawColor,line width= 0.4pt,dash pattern=on 4pt off 4pt ,line join=round,line cap=round] ( 49.20,111.78) -- (227.75,111.78);
\definecolor{drawColor}{RGB}{0,107,69}

\path[draw=drawColor,line width= 0.4pt,line join=round,line cap=round] ( 55.81,111.78) --
	( 55.81,154.40);
\end{scope}
\end{tikzpicture}}}
	\end{minipage}
	\caption{Auxiliary plots for explanation of the WM case: (a) breakdown of the space of Utility into four sectors and (b) buy and sell orders of Utility in the WM case}
	\label{UtilitySectorsAndWM}
\end{figure}

Much less intuitive is the functioning of the WM equilibrium.  The main characteristic of the wholesale market is an assumption that Utility is allowed to engage into speculation.\footnote{{Please note that by speculation we only assume that Utility tries to purchase electricity instead of producing it. The forthcoming description will show that this speculation is risk-free.}}    As a result, the bidding strategy of Utility changes. On the other hand, the behavior of both Retailer and Supplier remains the same.  This case is illustrated in Figure \ref{WM} and is supported by the lower part of Table \ref{TABLESUPPL}. Moreover, to ease the forthcoming explanation, we divided the Utility's supply and demand curves into 4 sectors. These sectors are illustrated in Figure \ref{UtilitySectors}.

\begin{table}[h!]
	\centering
	\begin{tabular}{|c|c|c|}
		\hline
		& Supply Pool & Demand Pool\\
		\hline
		\multirow{2}{24pt}{ASD}		& Sell orders of Supplier & Buy orders of Retailer\\
		& Sell orders of Utility & Buy orders of Utility \\ 
		\hline 
		\multirow{3}{24pt}{WM}		& Sell orders of Supplier & Buy orders of Retailer\\
		& Sell orders of Utility from Sector 1 & Buy orders of Utility from Sector 4 \\ 
		& Buy orders of Utility from Sector 2 & Sell orders of Utility from Sector 3 \\
		\hline  
	\end{tabular}
	\caption{Supply and Demand pools in the ASD and WM cases} 
	\label{TABLESUPPL}
\end{table}

Let us now consider the demand curve in the WM scenario. Given that Retailer can not act as a speculator, we place all its buy orders into the demand pool. Then, we shift our focus to Utility. We include orders located in sectors 3 and 4 in the demand pool. The following rationale can be used to justify the choice of these orders. First, the buy orders in sector 4 are located to the right of the internal equilibrium of Utility. Therefore, these orders are out-of-the-money and Utility can not capitalize on them. Utility thus hopes that they may be interesting to other market players. Hence, the buy orders from sector 4 end up in the demand pool. Second, placing the sell orders from sector 3 in the demand pool at first seems counterintuitive. However, we suppose that Utility decides to halt its own electricity production. Instead, Utility tries to purchase these orders at the same prices in the wholesale market. Purchasing rather than generating electricity may allow Utility to e.g. spare some maintenance costs.\footnote{Other possible reasons for speculation were mentioned in subsection \ref{InelastSubs}} Hence, Utility no longer tries to produce the orders located in sector 3 itself. On the contrary, Utility attempts to buy these orders in the market. Therefore, they are placed in the demand pool in the WM case.  

At this stage there are three groups of orders left unassigned. They will thus become components of the supply pool. More specifically, the three groups are: the sell orders of Supplier, the sell orders of Utility from sector 1, and the buy orders of Utility from sector 2. Let us now discuss why the orders from these groups are placed into the supply pool. Given the above discussion, the treatment of the former group appears straightforward. The latter two groups, in turn, require a special attention. Analogously to the Utility's buy orders from sector 4, the sell orders in sector 1 are out-of-the-money. Therefore, Utility attempts to find new customers among other market participants for these orders. Finally, intuition behind manipulations with the buy orders from sector 2 is connected to the sell orders from sector 3. From the previous paragraph we know that Utility intends to purchase the sell orders from sector 3 instead of producing them. To be able to carry out this transaction, Utility needs money. Thus, Utility places the buy orders from sector 2 in the supply pool. In doing so, Utility hopes that these orders would be acquired by other market participants. If other market agents indeed buy these orders, Utility obtains necessary funds to engage into the speculation and purchase the sell orders from sector 3. 

Figure \ref{UtilityWM} may be useful to amend the above explanation. This Figure shows us supply and demand schedules of Utility in the WM case. Let us now examine these schedules in greater detail. First, we can see explicitly that the downward-sloping demand curve incorporates the sell orders from sector 3. Second, the buy orders from sector 2 were included into the upward-sloping supply curve. Third, there are no orders in the downward-sloping demand curve which are more expensive than the internal equilibrium price of Utility. Fourth, there are no orders in the upward-sloping supply curve which are cheaper than the internal equilibrium price of Utility. Finally, the internal equilibrium price of Utility remains the same in both ASD and WM cases.

Having elaborated on the fundamentals of both ASD and WM  equilibria, we can now study their features. First of all, note that the equilibrium prices are the same in both cases. On the contrary, the equilibrium volumes are higher in the latter case. The discrepancy in the volume sizes can be explained by the fact that Utility tries to abstain from production of electricity in the WM scenario. Naturally, this assumption goes in line with subsection \ref{InelastSubs}. Therefore, market clears under an assumption that Utility does not supply some electricity to the market. Second, elasticity of the demand curve is typically higher in the WM case. This fact seems obvious given that Utility tries to speculate in the wholesale market and is thus more sensitive to price fluctuations. 

Moreover, there is a final commentary regarding the two equilibria. Note that Utility may realize that speculating is no longer profitable after the equilibrium price has been established. In other words, Utility will chose to refrain from speculation if its internal costs for electricity generation are lower than the market prices. In this case final load values in both ASD and WM scenarios will coincide. {From this perspective, Utility does not face any risks when trying to speculate in a wholesale market. Therefore, Utility's orders in the wholesale supply and demand curves are arbitrage orders. }

\subsection{Main idea and motivation} \label{SectionMainMaidea}
Following \cite{coulon2014hourly}, a model with perfectly inelastic demand is an extreme representation of an electricity market. Of course, transferring all demand elasticities to the supply side allows us to simplify computations substantially. However, sensitivity of some market agents to the price should not be neglected. Thus, the model with perfectly inelastic demand remains an imperfect solution. On the other hand, the wholesale market equilibrium, too, is not fully informative. The presence of arbitrage orders in this equilibrium distorts true intentions of market participants. Therefore, as the paper by \cite{knaut2016hourly} suggests, the ASD equilibrium can provide us with a more precise approximation of a wholesale electricity market. 

In fact, the ASD solution lies "in between" the solution with inelastic demand and the WM case. Therefore, the equilibrium volumes obtained in these two cases are respectively greater and lower than those in the ASD equilibrium. Moreover, the ASD market volumes, as we will later demonstrate, exhibit a stronger correlation with true total load values in an electricity market. Better suitability of the ASD solution for load approximation can be easily explained by commentaries made in the previous sections. 

\begin{figure}[h]
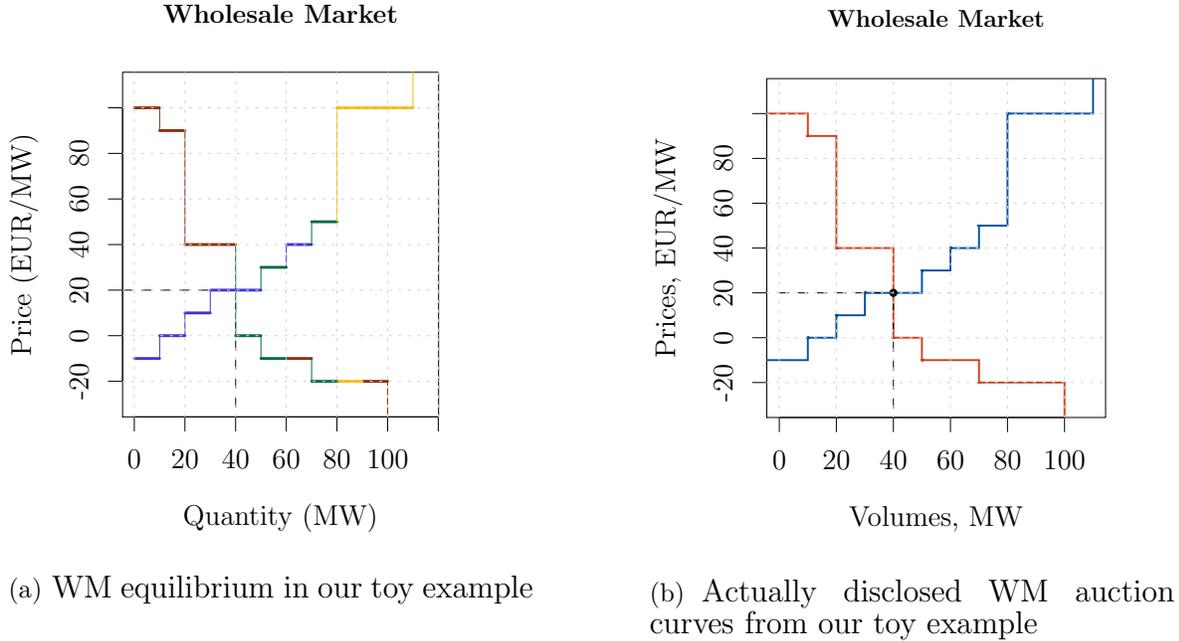

	\hspace*{1cm}
	\parbox{2.2in}{%
		\hspace{-0.5cm}\subfigure[\normalsize WM equilibrium in our toy example \label{WMAnother}]{\scalebox{0.9}{\input{DemGrSin.tex}}}}
	\begin{minipage}{01in}%
		\hspace*{2.3cm}\subfigure[\normalsize Actually disclosed WM auction curves from our toy example \label{WMReal}]{\scalebox{0.9}{\input{MarketRealToyBig.tex}}}
	\end{minipage}
	\caption{Auxiliary plots for explanation of the WM case: (a) wholesale market equilibrium in our toy example (b) auction curves in this equilibrium as are typically disclosed by an electricity exchange}
	\label{TwoWMEq}
\end{figure}

The model described in section \ref{ASDWM} may suggest that obtaining the ASD solution is relatively simple. In reality, however, it not necessarily is. Many electricity exchanges, including e.g. the German EPEX SPOT SE, do not disclose individual buy and sell orders of market participants. Therefore, the data available to scientists only constitutes auction curves drawn in a wholesale market. Precise compositions of these curves and contributions of each market player to these curves is not announced publicly. Hence, if we remain in the world of our toy example, the auction curves which are typically disclosed to market participants look like those in Figure \ref{WMReal} and not like those in Figure \ref{WMAnother}. 

Nevertheless, we have a certain knowledge about how exactly both ASD and WM solutions can be assembled. Given this knowledge, it is possible to derive the supply and demand schedules of market participants from the wholesale market data. More specifically, our optimization-based model will back-decompose the wholesale auction curves into individual buy and sell orders of market agents. The ASD equilibrium can be constructed easily once the decomposition is complete. 


\section{Model} \label{ModelMain}
\subsection{Intuition behind the functioning of the model} \label{ModelIntuition}

The to-be-undertaken decomposition of the wholesale auction curves is relatively sophisticated. Prior to explaining its details, two comments are to be made. First, in what follows we will use the name Fundamental Model (FM) to refer to the ASD case. The choice of the name appears rational because we will no longer apply our model to an imaginary data set. Therefore, our ASD solution will correspond to a fundamental model of a wholesale electricity market. Second, the decomposition will be formulated as an optimization problem. To obtain our FM model, we will minimize the distance between the FM equilibrium volumes and the true load values. In fact, the decomposition will be carried out along the five steps listed below. Referring to these steps may ease the understanding of the model's description. 

\begin{enumerate}
	\item Define the basic mathematical framework of the model
	\item Conjecture the internal equilibrium price of Utility. Since individual supply and demand schedules of market participants remain undisclosed, there is no information regarding the internal equilibrium of Utility. Hence, this step is critical because of two main reasons. First, the only difference between the ASD and WM cases lies in the Utility's bidding strategy. Second, the Utility's behavior is solely predetermined by its internal equilibrium price. Therefore, making an assumption regarding this price is necessary to perform the decomposition.
	\item Break down the wholesale supply and demand curves into individual supply and demand schedules of market participants. To carry out the decomposition, we will rely on the conjectured internal price of Utility and on our knowledge about the functioning of the wholesale market. Moreover, important is that at this stage we will obtain supply and demand curves of Utility in the WM setting.\footnote{The schedules of Utility in this setting were depicted in Figure \ref{UtilityWM}} These curves have to be transformed from WM to FM case in the next step. 
	\item Convert the Utility's curves from WM to FM setting.
	\item Combine the obtained supply and demand schedules of the market participants in an FM equilibrium. Ensure that the distance between the FM market volumes and true load values is minimized. 
\end{enumerate}

\subsection{Model description} \label{ModelDescription}
\subsubsection{Defining the basic mathematical framework of the model} \label{MathFrame}
To explain the model, we will still remain in the world of the toy example introduced in section \ref{ASDWM}. Before we proceed further, let $\pmin = -20$ and $\pmax=100$. These two borders correspond to the ones given in the toy example. We thus can define the supply and demand curves in a wholesale market  as follows
\begin{align}
\WSup:& \underbrace{(0, \infty)}_{\text{Volumes}} \mapsto [\pmin, \pmax]\label{WSsup}\\
\WDem:& (0, \infty) \mapsto [\pmin, \pmax]\\
\intertext{with the respective inverses being given by}
\WSupInv:&  [\pmin, \pmax] \mapsto (0, \infty) \label{WSupInv}\\
\WDemInv:& [\pmin, \pmax] \mapsto (0, \infty). \label{WDemInv}
\end{align}
Please note that the above defined curves correspond to those depicted in Figure \ref{WMReal}. This implies that all manipulations described below will be performed on auction curves with unknown compositions, i.e. on curves without a clear breakdown into individual orders of market participants. 

The market equilibrium can be clearly determined as follows
\begin{align}
(v^W, p^W) = \{v, \WSup(v) | \WSup(v)=\WDem(v), v>0\} \label{EQUILW}
\end{align}
where $v$ denotes the volume and $W$ stands for the wholesale market. 

\subsubsection{Conjecturing the internal equilibrium price of Utility}
As was mentioned earlier, the internal equilibrium price of Utility predetermines the differences between the ASD and WM cases. Since this price is typically not announced publicly, we need to conjecture it.  

Let us denote the internal equilibrium price of Utility by $p^U$. We suppose that price $p^U$ is a linear function of the wholesale market price. This assumption, despite being simple, appears reasonable. Utility would not be able to operate  in a market successfully if its prices would substantially deviate from market prices. The following statement can hence be made
\begin{align}
p^U = a_0 + a_1 p^W \label{UTPR}
\end{align}
where $a_0$ and $a_1$ denote the intercept and the slope, respectively. Finally, to ease the forthcoming description of the model, we suppose that the Utility's internal equilibrium price is greater than the respective equilibrium market price. 

\subsubsection{Breaking down the wholesale supply and demand curves into individual supply and demand schedules of market participants}
In this section we will determine supply and demand schedules of Supplier, Retailer, and Utility in the WM setting. As has already been said, the strategies of Supplier and Retailer do not vary between the FM and WM equilibria. The strategy of Utility, however, does. Hence, the two obtained supply and demand curves of Utility will be modified further in subsection \ref{ConvertingFMWM}. To be able to distinguish market participants from one another, will use index $0$ to denote Supplier or Retailer, and will use index $1$ for Utility.   

\vspace{0.2cm}
\noindent\textit{Step 1: determining the supply and demand schedules of Supplier and Retailer}
\vspace{0.2cm}

First of all, let us return to Figure \ref{WM} and focus on the wholesale supply curve. As can be seen from the Figure and as was mentioned in section \ref{ASDWM}, Utility will not place an order on the supply side if this order is priced below the Utility's internal equilibrium price $p^U$. Therefore, all orders in the wholesale supply curve which are cheaper than $p^U$ belong to Supplier. In turn, orders more expensive than the price $p^U$ can either be posted by Utility or by Supplier. In other words, the wholesale supply curve above the price $p^U$ is split in some proportion between Utility and Supplier.\footnote{We will calculate this proportion by using an optimization tool later.} 

The supply schedule of Supplier is thus built up of two components. The first component is the segment of the wholesale supply curve below price $p^U$. The second component is a part of the upper segment of the wholesale supply curve (above price $p^U$). The corresponding mathematical representation can be written as follows
\begin{align}\SupInv_0(p) &= \ione_\pminpu \WSupInv(p) \nonumber\\ &\qquad+\ione_\pupmax\left(\WSupInv(p^U) + (1-\gamma_1) \left(\WSupInv(p)-\WSupInv(p^U)\right)\right)\label{SUPSCHEDULE}\end{align}
where coefficient $\gamma_1$ denotes the Utility's proportion in the upper segment of the wholesale supply curve. 

Of course, the demand schedules of Retailer can be derived analogously. Following Figure \ref{WM}, the upper segment of the wholesale demand curve (above the price $p^U$) belongs solely to Retailer. The segment below price $p^U$ is split in some proportion between Retailer and Utility. Hence, it holds that 
\begin{align}\DemInv_0(p) &= \ione_\pupmax \WDemInv(p) \nonumber\\ 
&\qquad+\ione_\pminpu\left(\WDemInv(p^U) + (1-\phi_1) \left(\WDemInv(p)-\WDemInv(p^U)\right)\right) \label{DEMSCHEDULE}\end{align}
where coefficient $\phi_1$ denotes the Utility's proportion in the lower segment of the wholesale demand curve. 

\vspace{0.2cm}
\noindent\textit{Step 2: determining supply and demand schedules of Utility in the WM setting}
\vspace{0.2cm}

\begin{figure}
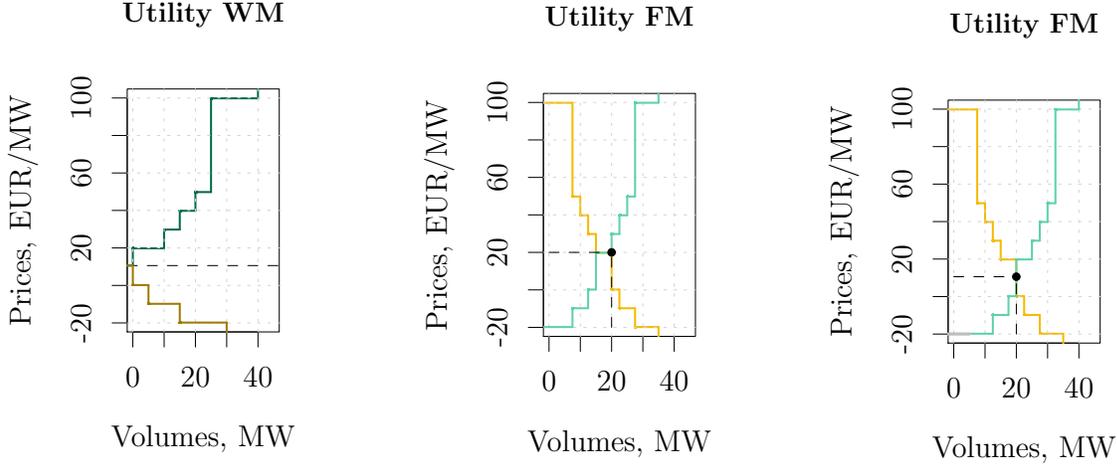

	\centering
	\parbox{2.4in}{%
		\vspace{-0.3cm}
		\subfigure[\normalsize Wholesale auction curves of Utility with $\gamma_1=0.5$ and $\phi_1=0.5$ \label{UtilityRealToyBig}]{\scalebox{0.95}{\input{UtilityRealToyBig.tex}}}}
	\begin{minipage}{2in}%
		\hspace{-0.7cm}\subfigure[\normalsize Unadjusted curves of Utility in FM setting with $\alpha_1=0.5$ and $\beta_1=0.5$ \label{UtilityRearrangedToyBig1}]{\scalebox{0.95}{\input{UtilityRearrangedToyBig1.tex}}}
	\end{minipage}%
	\begin{minipage}{2in}%
		\hspace*{-0.4cm}\subfigure[\normalsize Adjusted curves of Utility in FM setting with $\alpha_1=0.5$ and $\beta_1=0.5$ \label{UtilityRearrangedToyBig2}]{\scalebox{0.95}{\input{UtilityRearrangedToyBig2.tex}}}
	\end{minipage}%
	\caption{Auxiliary plot for the explanation of the WM case: transformation of the Utility's curves from WM to FM setting}
	\label{UtilityCurvesTransformation}
\end{figure}

From equation \ref{SUPSCHEDULE} and Figure \ref{UtilityWM} we know that the upper segment of the wholesale supply curve  is split between Supplier and Utility. Therefore, the upward-sloping supply curve of Utility in the WM case can be written as
\begin{align}\WSupInv_1(p) = \ione_\pupmax \gamma_1\left(\WSupInv(p)-\WSupInv(p^U)\right). \label{WSup1} \end{align}
In turn, following equation \ref{DEMSCHEDULE}, the downward-sloping curve can be represented as
\begin{align} 
\WDemInv_1(p) = \ione_\pminpu \phi_1\left(\WDemInv(p)-\WDemInv(p^U)\right). \label{WDem1}
\end{align}
Hence, equations \ref{WSup1} and \ref{WDem1} define Utility's auction curves in the WM case as depicted in Figure \ref{UtilityRealToyBig}. The following section will be devoted to transforming these curves from WM to FM setting.

\subsubsection{Converting the Utility's schedules from wholesale market to fundamental model setting} \label{ConvertingFMWM}
As section \ref{ASDWM} suggests, Figure \ref{UtilityRealToyBig} corresponds to Figure \ref{UtilityWM}. However, to be able to construct the FM equilibrium, Utility's auction curves in the initial setting (as depicted in e.g. Figure \ref{UtilitySectors} ) are required. Transforming the supply and demand curves of Utility from WM to FM case is thus the main aim of the present subsection. This transformation will be carried out in two steps.

\vspace{0.2cm}
\noindent\textit{Step 1: separating the Utility's WM curves into buy and sell orders}
\vspace{0.2cm}

Remember that the supply and demand curves depicted in Figure \ref{UtilityWM} incorporate both buy and sell orders. However, following section \ref{SectionMainMaidea} and Figure \ref{UtilityRealToyBig}, we do not know the exact compositions of the Utility's curves in our case. Therefore, we have to divide the Utility's WM curves into buy and sell orders. Following the general structure of the model, an optimization-based technique will be used to determine the proportions in which orders are to be divided. 

Let us now consider the downward-sloping demand curve in Figure \ref{UtilityRealToyBig}. Suppose that orders in this curve are split equally between the buy and sell sides. To quantify this assumption, we introduce a new coefficient $\alpha_1=0.5$. Then, following Figure \ref{UtilitySectorsAndWM}, half of the downward-sloping demand curve remains on the buy side. On the contrary, the other half of this curve is flipped onto the supply side. In other words, as we can see from Figure \ref{UtilityRearrangedToyBig1}, the lower part (below price $p^U$) of the supply curve of Utility FM is a flipped half of the downward-sloping demand curve of Utility WM. Naturally, the supply and demand curves below the price $p^U$ in the Utility FM case look like mirrored versions of one another because $\alpha_1=0.5$. 

We can now use the same method to decompose the supply curve of Utility WM. Let us thus focus on the upward-sloping curve in Figure \ref{UtilityRealToyBig}. Assume that the orders in this curve, too, are split equally between the buy and sell sides. Therefore, let $\beta_1=0.5$. As the description provided in the previous paragraph suggests, one half of the upward-sloping curve is flipped onto the demand side. Hence, as can be seen from Figure \ref{UtilityRearrangedToyBig1}, the upper part of the demand curve (above the price $p^U$) in the FM case is a flipped half of the upward-sloping supply curve of Utility WM. 

Hence, the supply curve depicted in Figure \ref{UtilityRearrangedToyBig1} can be represented mathematically as
\begin{align}
\wtilde{\FSup}^{-1}_1(p) &= \ione_\pminpu \alpha_1\left(\WDemInv_1(\pmin)-\WDemInv_1(p) \right) \nonumber\\
& \qquad + \ione_\pupmax \left(\alpha_1 \WDemInv_1(\pmin) + (1-\beta_1) \WSupInv_1(p)\right) \label{FSup1Prel}
\end{align}
where $\alpha_1$ is a portion of sell orders in the downward-sloping curve of Utility WM and $\beta_1$ denotes a portion of buy orders in the upward-sloping curve of Utility WM. The first line of the above equation is thus a horizontally flipped part of the demand curve of Utility WM.\footnote{Note that the assumptions $\alpha_1=0.5$ and $\beta_1$ were only made to simplify the description of the model. The actual values of $\alpha_1$ and $\beta_1$ have to be determined by means of an optimization tool.}

In turn, the definition of the demand curve illustrated in Figure \ref{UtilityRearrangedToyBig1} reads
\begin{align}
\wtilde{\FDem}^{-1}_1(p) &= \ione_\pupmax \beta_1\left(\WSupInv_1(\pmax)-\WSupInv_1(p) \right) \nonumber\\
& \qquad + \ione_\pminpu \left(\beta_1 \WSupInv_1(\pmax) + (1-\alpha_1) \WDemInv_1(p)\right). \label{FDem1Prel}
\end{align} 
Equations \ref{FSup1Prel} and \ref{FDem1Prel}, however, do not describe the final versions of Utility's curves in the FM case. These equations will be modified further in the next step.

\vspace{0.2cm}
\noindent\textit{{Step 2: reconciling equilibrium prices of Utility WM and Utility FM}}
\vspace{0.2cm}

From section \ref{ASDWM} and Figure \ref{UtilitySectorsAndWM} we know that the internal equilibrium price of Utility is the same in WM and FM settings. On the other hand, Figures \ref{UtilityRealToyBig} and \ref{UtilityRearrangedToyBig1} indicate explicitly that prices may differ after the decomposition of the wholesale auction curves. Naturally, this discrepancy violates our assumption. Therefore, we have to further manipulate the curves of Utility FM for the equilibrium prices of Utility in both WM and FM settings to be reconciled. 

{Our solution to solve the problem is rather simple.  We merely decide to shift one of the Utility's curves to the right unless equilibrium prices in FM and WM cases are reconciled.\footnote{Intuition behind which curve needs to be shifted will be elaborated in what follows.} Despite being counterintuitive at first, this approach proves to be extremely efficient and does not contradict to the main aim of our model. Figures \ref{UtilityRearrangedToyBig1} and \ref{UtilityRearrangedToyBig2} provide us with the corresponding graphical representations.  

Of course, shifting one of the curves to the right may imply that we "create" additional volumes. One would thus expect that we should subtract the added volumes from other parts of the curves. However, when the model is applied to real data, subtracting these volumes will inevitably distort either the internal price of Utility, or the equivalence of the prices in final WM and FM equilibria. Therefore, a question arises: where do these additional volumes come from? 

Let us now suppose that we shift the Utility's supply curve.\footnote{Of course, the reasoning is analogous for the demand side.} Note that the need to maintain the stability of an energy system forces energy producers to offer must-run supply at minimal prices. Therefore, sell orders at the lowest price are typically much more voluminous than other sell orders. Following section \ref{ASDWM}, these must-run sell orders are located  in the segment $\pmin$ of the downward-sloping demand curve of Utility WM. 
Hence, when converting the schedules of Utility form WM to FM setting, we, theoretically, should not only shift the supply curve of Utility FM to the right. We should also subtract the shift size from the segment $\pmin$ of the demand curve of Utility FM. 

However, important is the fact that this segment is represented graphically only in the toy example. It is not displayed {in real-world examples}. Actually observed wholesale auction curves almost abruptly end at the prices $\pmin$ and $\pmax$.\footnote{Figure \ref{2Graphs} plots two auction curves observed in the German day-ahead wholesale electricity market.} In other words, we indeed can subtract the volumes in question, but only if we remain in the world of the toy example. Moreover, it is worthy to note that manipulations with the far-right segments of the curves influence neither Utility's price nor the prices in WM and FM cases. The equilibrium remains unchanged even if no volumes are subtracted. The shape of the demand curve, too, remains unaffected.}




Figure \ref{LoadGraph} will provide an even greater motivation for using the second approach. As we will later demonstrate, equilibrium volumes we obtain approximate the actual load values better than the model with perfectly inelastic demand and the wholesale market volumes. Therefore, after several adjustments our model can be used successfully for load forecasting. 

Let us now explain how the shift of Utility's curves is implemented into the model. First, we will denote the magnitude of the shift by $\tau_1$ and assume that
\begin{align}
\tau_1 = \wtilde{FSup}_1^{-1}(p^U) - \wtilde{FDem}_1^{-1}(p^U) \label{TauShift}
\end{align}
Hence, we determine the shift size by calculating the difference between volumes on supply and demand curves at point $p^U$. From equation \ref{TauShift} it appears clear that $\tau_1$ is negative if the demand curve is located to the right of the supply curve at the point $p^U$. Hence, we will shift the Utility's supply curve if $\tau_1<0$. Otherwise, if $\tau_1>0$, then the demand curve has to be shifted. 

Then, the equations for the adjusted supply and demand curves of Utility FM can be respectively represented as
\begin{align}
{\FSupInv}_1(p) &= \ione_\pminpu \alpha_1\left(\WDemInv_1(\pmin)-\WDemInv_1(p) \right) \nonumber\\
& \qquad +\ione_\pupmax \left(\alpha_1 \WDemInv_1(\pmin) + (1-\beta_1) \WSupInv_1(p)\right) - \min(\tau_1,0) \label{FSup1}
\end{align}
and 
\begin{align}
{\FDemInv}_1(p) &= \ione_\pupmax \beta_1\left(\WSupInv_1(\pmax)-\WSupInv_1(p) \right) \nonumber\\
& \qquad +\ione_\pminpu \left(\beta_1 \WSupInv_1(\pmax) + (1-\alpha_1) \WDemInv_1(p)\right) + \max(\tau_1,0). \label{FDem1}
\end{align} 

Figure \ref{UtilityRearrangedToyBig2} plots the final curves of Utility in the FM case. As can be seen from the Figure, the equilibrium price of Utility FM equals to that of Utility WM once the adjustment was made. Moreover, in our case factor $\tau_1$ was positive since the initial supply curve of Utility FM is located to the left of the demand curve at the point $p^U$. Therefore, the supply curve was shifted to the right. The shift magnitude is highlighted in gray in Figure \ref{UtilityRearrangedToyBig2}.



\subsubsection{Assembling the Fundamental Model and formulating the optimization problem}

The above defined supply and demand schedules of the market participants allow us to build up the FM equilibrium. Hence, the market demand curve in the FM case can be obtained by aggregating the sell orders of Retailer and Utility. The following statement can thus be made
\begin{align}
\FDemInv=\DemInv_0+\FDemInv_1. \label{FDemFin}\\
\intertext{The market supply curve in the FM case can then be defined analogously as}
\FSupInv=\SupInv_0+\FSupInv_1 \label{FSupFin}
\end{align}

It follows that the intersection between the FM supply and demand curves can be defined as 
\begin{align}
(v^F, p^F) = \{(v, \FSup(v) | \FSup(v)=\FDem(v)), v>0\},
\end{align}
where $p^W=p^F$ and $v^W<v^F$ as follows from section \ref{ASDWM} and equation \ref{EQUILW}. 

As the previous sections suggest, values of $v^W$ in the wholesale market are much smaller than the actual load values. Note that not only arbitrage orders  influence the difference. Final load reflects volumes from intra-day trading (both the intra-day auctions and continuous trading), OTC-deals, and potentially other markets.\footnote{For example, the EXAA can be considered as an "other market" for the case of Germany.} Therefore, volumes $v^F$ in our FM equilibrium, too, must be smaller than the actual load values. Moreover, volumes $v^F$ must be lower than those suggested by the model with perfectly inelastic demand. This holds because the latter model leaves no ealsticities at the demand side.  

Finally, the core of our optimization model can be explained as follows. We will minimize the distance between the actual load values and the equilibrium volumes in the FM case. Speaking technically, a linear fit optimization problem can thus be written as follows
\begin{align}
\argmin_{\theta_0, \theta_1} Q(\theta_0, \theta_1) \hspace{0.5cm} \text{where} \hspace{0.5cm} Q(\theta_0, \theta_1) = \sum_{i=1}^{n} (load_i-\theta_0- \theta_1 v^F)^2\label{OPTPR}\end{align}


\subsection{Applying the model to the toy example} \label{ModelToyTest} 
\begin{figure}
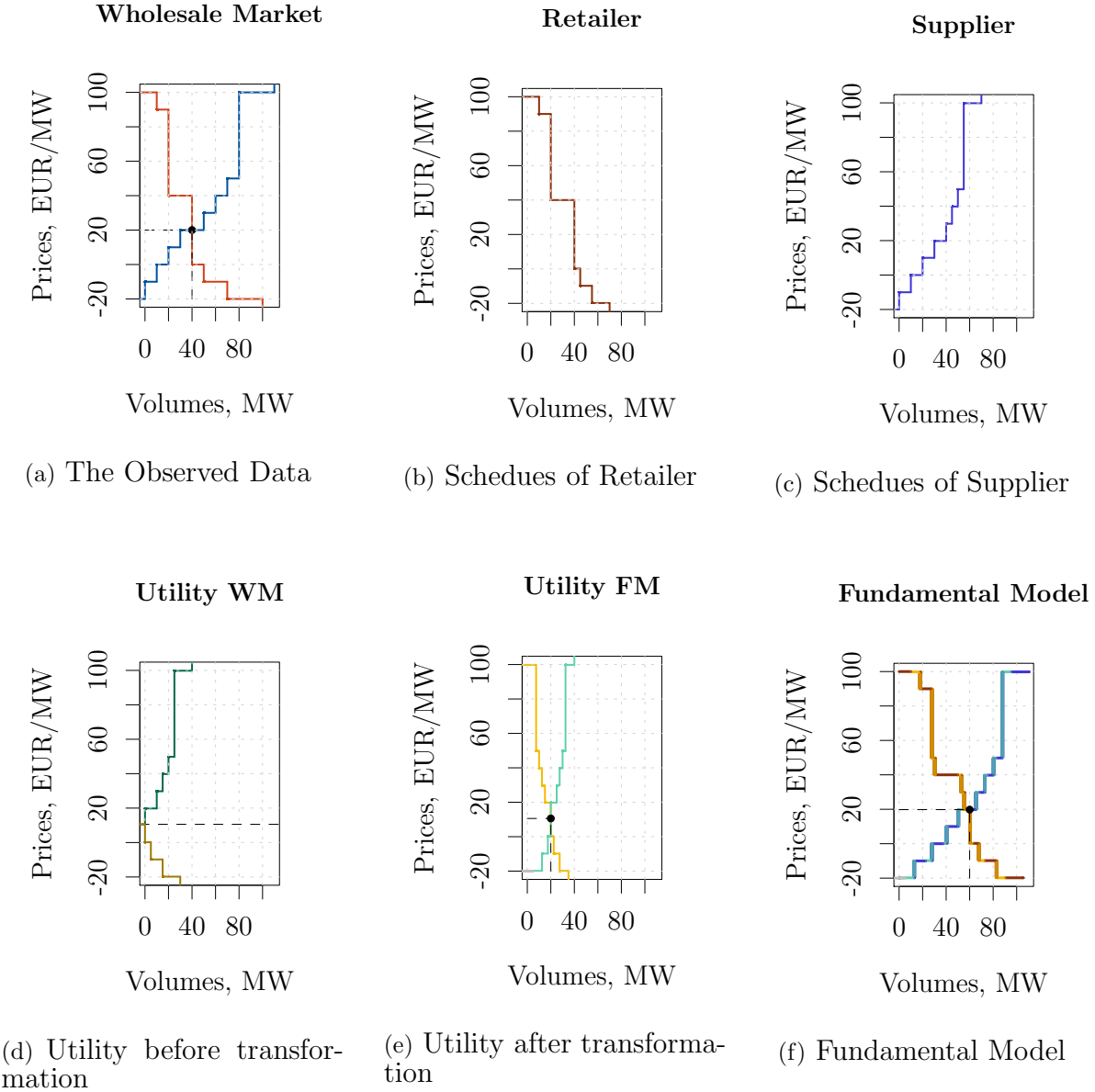

	\centering
	\parbox{2.4in}{%
		\vspace{-0.3cm}
		\subfigure[\normalsize The Observed Data  \label{WholesaleRealFigToy}]{\scalebox{0.95}{\input{MarketRealToy.tex}}}}
	\begin{minipage}{2in}%
		\hspace{-0.7cm}\subfigure[\normalsize Schedues of Retailer \label{RetailerRealFigToy}]{\scalebox{0.95}{\input{RetailerRealToy.tex}}}
	\end{minipage}%
	\begin{minipage}{2in}%
		\hspace*{-0.4cm}\subfigure[\normalsize Schedues of Supplier \label{SupplierRealToy}]{\scalebox{0.95}{\input{SupplierRealToy.tex}}}
	\end{minipage}%
	
	\parbox{2.4in}{%
		\subfigure[\normalsize Utility before transformation \label{UtilityRealToy}]{\scalebox{0.95}{\input{UtilityRealToy.tex}}}}
	\begin{minipage}{2in}%
		\hspace{-0.7cm}\subfigure[\normalsize  Utility after transformation \label{UtilityRearrangedToy}]{\scalebox{0.95}{\input{UtilityRearrangedToy.tex}}}
	\end{minipage}%
	\begin{minipage}{2in}%
		\vspace{-0.3cm}
		\hspace*{-0.4cm}\subfigure[\normalsize Fundamental Model \label{FMToy}]{\scalebox{0.95}{\input{FundamentalMarketToy.tex}}}
	\end{minipage}%
	\caption{Obtaining the Fundamental Model using the wholesale market data from the toy example with $a_0=0.5$, $a_1=0.5$, $\gamma_1=0.5$, $\phi_1=0.5$, $\alpha_1=0.5$, $\beta_1=0.5$}
	\label{FUNDMARKToy}
\end{figure}

{To test the functioning of our model, we first remain in the world of our toy example. Naturally, we use Figure \ref{WMReal} as the starting point. For illustration purposes we suppose the following set of coefficients: $a_0=0.5$, $a_1=0.5$, $\gamma_1=0.5$, $\phi_1=0.5$, $\alpha_1=0.5$, $\beta_1=0.5$.  The obtained model is presented in Figure \ref{FUNDMARKToy}.  There are two properties of our model to be noted. 

First, our knowledge of the bidding strategies allows us to compute the difference $v^F-v^W$ on the grounds of our model. From Figures \ref{ASD} and \ref{WM} we know that orders to the left of the equilibrium in the WM case can belong only to Retailer or Supplier. On the contrary, orders to the left of the equilibrium in the FM case can be submitted by Retailer, Supplier, and Utility. Therefore, the difference $v^F-v^W$ stems from the Utility's orders. The following statement can thus be made
\begin{align}
v^F-v^W=\begin{cases}
\FSupInv_1(p^U) -(\FDemInv_1(p^F)-\FDemInv_1(p^U))  \hspace{0.5cm} &\text{if} \hspace{0.5cm} p^U \leq p^F\\
\FDemInv_1(p^U) -(\FSupInv_1(p^F)-\FSupInv_1(p^U)) \hspace{0.5cm} &\text{if} \hspace{0.5cm} p^U \geq p^F \\
\end{cases}
\end{align}
Therefore, the relation between the prices $p^U$ and $p^F$ determines which of the auction curves has to be shifted to the right. Moreover, to compute the difference $v^F-v^W$ we have to focus on a curve that has not been adjusted. 

}

{Second, the model with perfectly inelastic demand is an extreme form of our model. Under the parameter setting $a_0=0$, $a_1=1$, $\gamma_1=1$, $\phi_1=1$, $\alpha_1=0$, $\beta_1=0$ all demand elasticities are transferred to the supply side.\footnote{{Implications of each of the coefficients are be elaborated at length in section \ref{ObtainedResults}}} More specifically, this setting first assumes that Utility's internal price equals to the wholesale market price. Second, Utility attains all possible orders from the wholesale auction curves. Third, when converting Utility's schedules from WM to FM case, all orders are transferred from the demand to the supply side. Therefore, the demand curve is represented only by one point, all volumes are accumulated on the supply side, and the final equilibrium volume reaches its maximum.  }

\section{Empirical Data} \label{Data} 
\subsection{Institutional framework of the German electricity market}
We chose to test our model on the German data. Wholesale electricity trading in Germany takes place on several different marketplaces. Their timing is well illustrated in \cite{kiesel2017econometric}. Their detailed overview is provided on the EPEX SPOT SE website or in e.g. \cite{hagemann2013empirical}. Additionally to wholesale markets, electricity is actively traded via various OTC-contracts. 

First of all, a day-ahead auction for hourly products is conducted at 12:00 on a daily basis. Following the rules of the exchange, minimal price in the day-ahead market equals to $\pmin=-500$ EUR and maximal price is limited to $\pmax=3000$ EUR. Market participants are expected to submit their bids to an auctioneer prior to the start of the auction. Once the bidding window closes, the EPEX system matches the orders, constructs wholesale supply and demand auction curves, and establishes 24 hourly prices for the next day. These prices are announced as soon as possible from 12:42.  Besides the prices, also the auction curves are disclosed publicly. However, these curves are revealed only in their aggregated form, i.e. individual orders of market participants still remain concealed. Moreover, there exists another day-ahead auction for 15-minutes contracts. This auction clears at 15:00 every day. 

In addition to the day-ahead market, a continuous intra-day trading for hourly contracts begins at 15:00 the day before physical electricity delivery and closes 30 minutes prior to the delivery. Continuous auction for 15-minutes contracts begins at 16:00 daily. Details of the German continuous intra-day market are discussed at length in e.g. \cite{von2017optimal}. 

Finally, energy can be traded in a balancing market to eliminate discrepancies between electricity supply and demand. Imbalances are traded ex-post. This market is studied at length in e.g. \cite{just2015strategic}.

\subsection{Data Set} 
Our sample period extends from $31.12.2016$ to $31.12.2017$. To decrease computational burden of our model, we will set the upper and lower price bounds to $p_{\min, 2017}=-83.05$ EUR and $p_{\max, 2017}=163.50$ EUR. The selected values correspond to the minimum and maximum observation, respectively, present in the in-sample period. 

\begin{figure}[h]
	\centering
	\vspace{-1cm}
	\scalebox{0.9}{\input{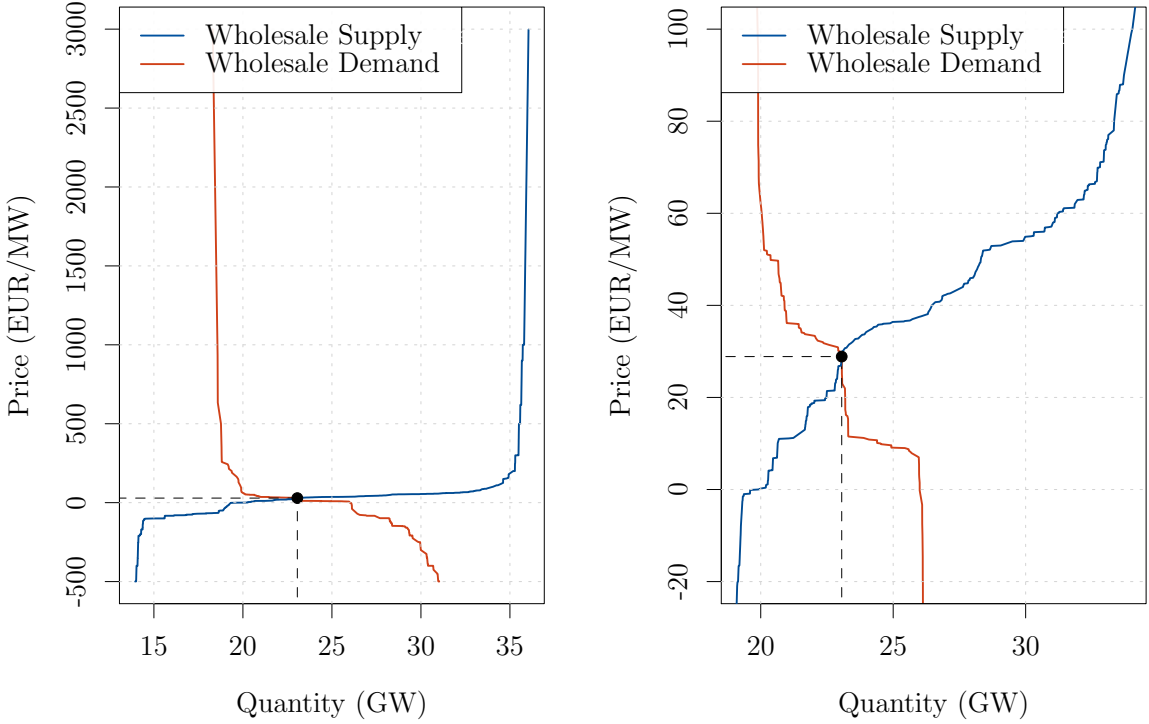}}		
	\caption{Wholesale market equilibrium on 2017-01-01 23:00:00. The left plot shows the entire auction curves observed at the time point. The right plot focuses on the equilibrium between these two curves.}
	\label{2Graphs}
\end{figure}

An example of the collected auction curves data is provided in Figure \ref{2Graphs}.

{In fact, to define wholesale supply and demand curves (equations \ref{WSupInv} and \ref{WDemInv}), we used a non-equidistant price grid with the following specification
\begin{align*}
pg=\{-500,-450,-400,...,-83.05,-82.85,-82.65, ... , 163.50, 213.50, 263.50,...,3000\},
\end{align*}
where $pg$ stands for price grid. The use of such price grid allows us to incorporate all segments of the supply and demand curves while retaining a particular focus on their most important parts, i.e. on the interval from $p_{\min,2017}$ to $p_{\max,2017}$. Therefore, the computational burden of our model lessens, though the quality of the results remains unchanged. }

Additionally to the auction curves data, we used total load data obtained from the ENTSOE. This data was though provided in a quarter-hourly format. Given that the auction curves data had a hourly resolution, we used simple arithmetic averages to manipulate the load data. Moreover, all data sets were clock-change adjusted. To replace missing hours in March, we used two values before and after these hours. In turn, simple arithmetic averages of two double hours in October were calculated to solve the problem. 

%


\section{Results} \label{Results} 
\subsection{The obtained coefficients}\label{ObtainedResults}
The functioning of our model on real data is depicted in Figure \ref{FUNDMARK} below. This Figure is based on {365} in-sample observations with $31.12.2016$ being the starting point. The obtained coefficients for the year 2017 are summarized in Table \ref{TAB1} below. {The conducted robustness checks show that the coefficients are significantly different from zero. A weak significance level is exhibited only by coefficient $\beta_1=0.019$. Moreover, all but coefficient $\phi_1=0.984$ are significantly different from one. These results allow us to conclude that the assumptions on which our model is built are valid and reasonable. }

\begin{table}[h!]
	\centering
	\begin{tabular}{|c|c|c|c|c|c|}
		\hline
		$a_0$ & $a_1$ & $\gamma_1$ & $\phi_1$ & $\alpha_1$ &$\beta_1$\\
		\hline
		$\underset{(0.076)\textcolor{white}{eee}}{5.890_{\star\star\star}^{***}}$ & 
		$\underset{(0.006)\textcolor{white}{eee}}{0.963_{\star\star\star}^{***}}$ & 
		$\underset{(0.015)\textcolor{white}{eee}}{0.510_{\star\star\star}^{***}}$ & 
		$\underset{(0.008)\textcolor{white}{eee}}{0.984_{\circ}^{***}}$ & $\underset{(0.008)\textcolor{white}{eee}}{0.287_{\star\star\star}^{***}}$ & $\underset{(0.011)\textcolor{white}{eee}}{0.019_{\star\star\star}^{*}}$\\
		\hline
	\end{tabular}
	\caption{{The obtained coefficients for the year 2017 with the following significance levels: $\bullet=10\%$, $*=5\%$, $**=1$, $***=0.1\%$ with respect to 0; $\circ=10\%$, $\star=5\%$, $\star\star=1\%$, $\star\star\star=0.1\%$ with respect to 1.}}
	\label{TAB1}
\end{table}

Let us now analyze the above listed values more thoroughly. First, recall that coefficients $a_0$ and $a_1$ stand for the slope and intercept, respectively, in the linear function of the internal equilibrium price of Utility (equation \ref{UTPR}). The obtained values indicate explicitly that the internal price of Utility does not deviate substantially from the market price. Therefore, an earlier assumption can be confirmed, i.e. Utility would struggle to make profits if its strategy would be much different to that of other market players. 

The second pair of coefficients is $\gamma_1$ and $\phi_1$. They describe how the wholesale auction curves are split between the three market participants. Following equations \ref{SUPSCHEDULE} and \ref{WSup1}, the lower part of the wholesale supply curve (below the Utility's internal price $p^U$) belongs solely to Supplier. In turn, the upper part (above the price $p^U$) includes orders of both Utility and Supplier. As the value of $\gamma_1$ suggests, orders in the upper part of the wholesale supply curve are split almost equally between Utility and Supplier. Therefore, inelastic parts are present neither in the schedules of Supplier (Figure \ref{SupplierRealFig}) nor in the upward-sloping curve of Utility WM (Figure \ref{UtilityRealFig}). 

On the contrary, the split of the wholesale demand curve between Utility and Retailers is not equal. As was defined in equations \ref{DEMSCHEDULE} and \ref{WDem1}, the upper part of the wholesale demand curve (above the price $p^U$) contains only Retailer's orders, whereas orders of both Utility and Retailer can be incorporated in its lower part. The value of $\phi_1$ thus shows that Retailer retains only $0.016\%$ of the orders from the lower part of the wholesale demand curve. The remaining $0.984\%$ belongs to Utility. As a result, the lower part of the Retailer's demand curve is almost perfectly inelastic (Figure \ref{RetailerRealFig}).

Let us now shift our focus to coefficients $\alpha_1$ and $\beta_1$. These coefficients, as is given in equations \ref{FSup1} and \ref{FDem1}, determine the transformation of the Utility's schedules from WM to FM setting. The value of $\alpha_1$ implies that  roughly 30\% of the upward-sloping supply curve of Utility WM is flipped onto the demand side. On the other hand, the value of $\beta_1$ indicates that more than 98\% of the downward-sloping demand curve is transferred onto the supply side. Therefore, the lower part of the demand curve of Utility FM is almost perfectly inelastic. Moreover, the size of the lower part of the supply curve of Utility FM is almost maximal. Not only does Utility accumulate many orders from the wholesale demand curve (since $\phi_1$ is very small), but also Utility flips almost the entirety of those orders onto the supply side. 

Finally, the obtained supply and demand schedules of the market participants allow us to assemble the Fundamental Model. Note that the lower part of the demand curve of Utility FM is almost fully inelastic. The lower part of the Retailer's curve, too, is inelastic. Therefore, the demand curve in our Fundamental Model consists of two parts: a relatively elastic upper part and almost fully inelastic lower part. Thus, following our initial hypothesis, the demand curve in our Fundamental Model lies "in between" a perfectly inelastic one (derived in \cite{coulon2014hourly}) and the initial one in the wholesale market. 
	
Moreover, as has been mentioned earlier, the coefficient setting $a_0=0$, $a_1=1$, $\gamma_1=1$, $\phi_1=1$, $\alpha_1=0$ and $\beta_1=0$ yields the model with perfectly inelastic demand. Let us scrutinize this setting more thoroughly. The internal equilibrium price of Utility is equal to the wholesale market price  because $a_0=0$ and $a_1=1$. Then, Utility accumulates all possible sale and purchase orders from the wholesale supply and demand curves since $\gamma_1=1$ and $\phi_1=1$. $\alpha_1=0$ shows that the upward-sloping curve of Utility WM remains unmodified, whereas $\beta_1=0$ shows that the entire downward-sloping curve is flipped onto the supply side. As a result, equilibrium volumes are equal to those suggested by the model in \cite{coulon2014hourly}.

\begin{figure}
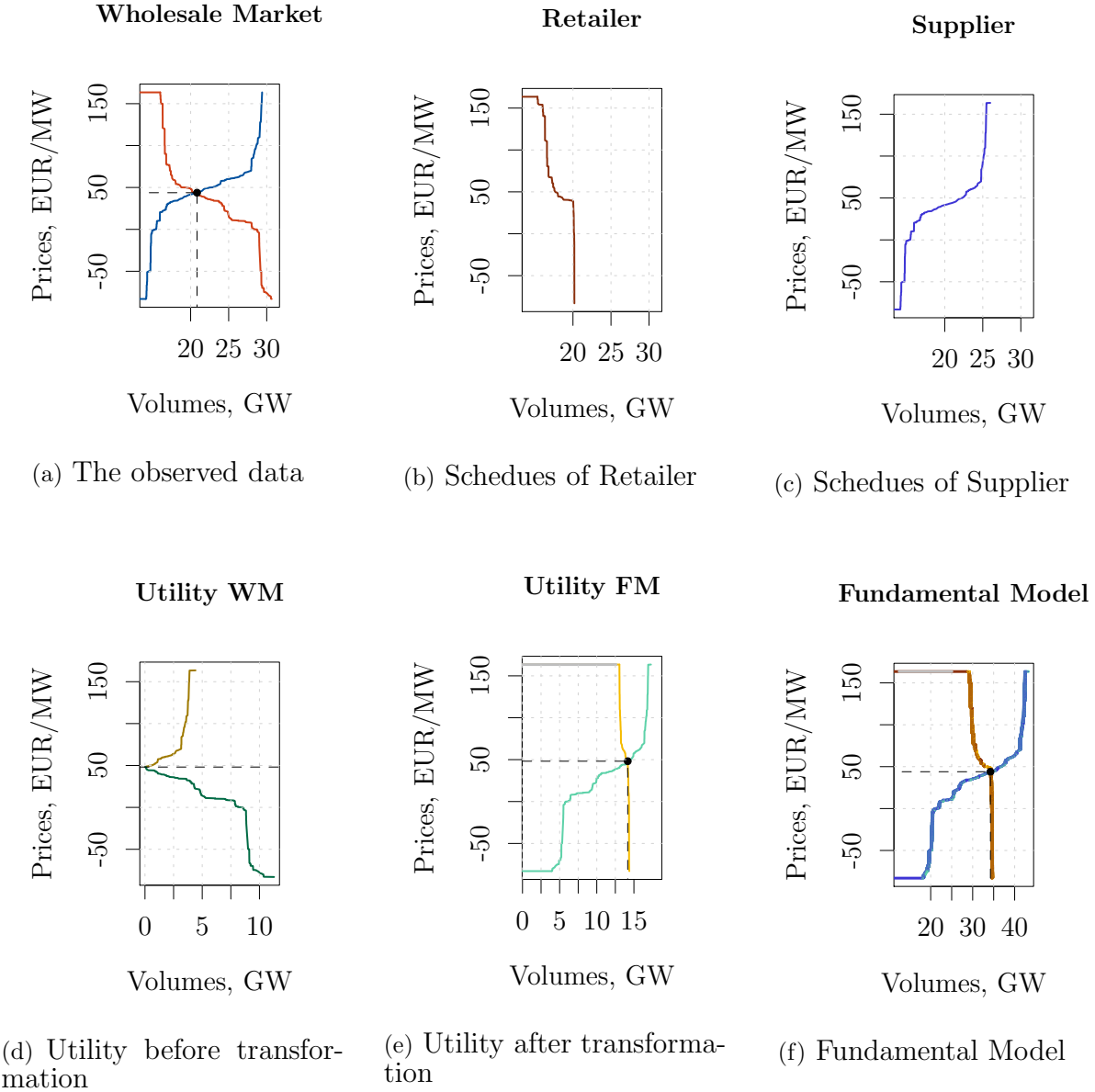

	\centering
	\parbox{2.4in}{%
		\vspace{-0.3cm}
		\subfigure[\normalsize The observed data  \label{WholesaleRealFig}]{\scalebox{0.95}{\input{MarketReal.tex}}}}
	\begin{minipage}{2in}%
		\hspace{-0.7cm}\subfigure[\normalsize Schedues of Retailer \label{RetailerRealFig}]{\scalebox{0.95}{\input{RetailerReal.tex}}}
	\end{minipage}%
	\begin{minipage}{2in}%
		\hspace*{-0.4cm}\subfigure[\normalsize Schedues of Supplier \label{SupplierRealFig}]{\scalebox{0.95}{\input{SupplierReal.tex}}}
	\end{minipage}%
	
	\parbox{2.4in}{%
		\subfigure[\normalsize Utility before transformation \label{UtilityRealFig}]{\scalebox{0.95}{\input{UtilityReal.tex}}}}
	\begin{minipage}{2in}%
		\hspace{-0.7cm}\subfigure[\normalsize  Utility after transformation \label{UtilityRearranged}]{\scalebox{0.95}{\input{UtilityRearranged.tex}}}
	\end{minipage}%
	\begin{minipage}{2in}%
		\vspace{-0.3cm}
		\hspace*{-0.4cm}\subfigure[\normalsize Fundamental Model \label{FM}]{\scalebox{0.95}{\input{FundamentalMarket.tex}}}
	\end{minipage}%
	\caption{Obtaining the Fundamental Model using the German wholesale market data on 2017-01-06 02:00:00 with $a_0=5.890$, $a_1=0.963$,  $\gamma_1=0.510$,  $\phi_1=0.984$, $\alpha_1=0.287$ and $\beta_1=0.019$}
	\label{FUNDMARK}
\end{figure}

\subsection{Actual electricity load and the values of $v^F$}
\begin{figure}[h!]
	\centering
	\vspace{-1cm}
	\resizebox{\textwidth}{!}{\input{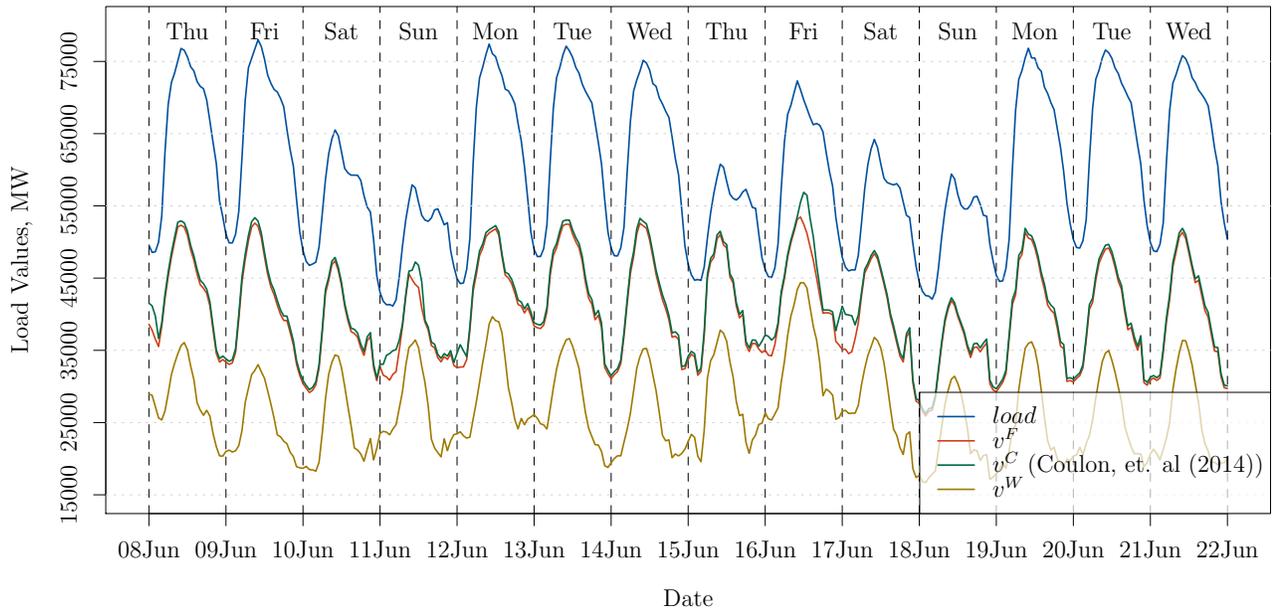}}		
	\caption{Fundamental Model volumes vs. values suggested by the \cite{coulon2014hourly} model vs. actual load values vs. wholesale market volumes for a two weeks sample from 08 June 2017 to 21 June 2017}
	\label{LoadGraph}
\end{figure}

{Besides the developed market decomposition model, the utility of our research can be well demonstrated by Figure \ref{LoadGraph}.  This Figure illustrates actual load values, wholesale market equilibrium volumes, Fundamental Model equilibrium volumes and volumes suggested by the model with perfectly inelastic demand for a two week sample 08-21 June 2017. To ease further notation, let us denote the volumes suggested by the \cite{coulon2014hourly} model by $v^C$.

From Figure \ref{LoadGraph} it can be seen clearly that the values produced by our Fundamental Model are often simply replicating the values of $v^C$. This observation seems intuitive given the fact that we often shift one of the Utility's curves as described in section \ref{ConvertingFMWM}. Volumes we add to reconcile Utility's prices push final values $v^F$ up. However, $v^F \leq v^C$ always holds since equilibrium volumes reach their maximum in the model with perfectly inelastic demand. Second, values $v^F$ have higher correlation with the actual load values than those of the model by \cite{coulon2014hourly}. To be more precise, the correlation between the load and $v^W$ equals to 0.36, between the load and $v^C$ to 0.63, and between the load and $v^F$ to 0.75. Closer correlation between $v^F$ and the load values can be seen explicitly during  e.g. night hours on 11th, 12th, 16th and 17th of June or during day hours on 11th and 16th of June. For instance, during the early morning on 11th of June the actual load values drop as is suggested by our Fundamental model, yet a contrary tendency is predicted by the\cite{coulon2014hourly} model. 

Furthermore, note that the red curve in Figure \ref{LoadGraph} can be easily shifted upwards to match the actual load values. 
In fact, shifting the red curve upwards eliminates the limitations imposed by earlier manipulations with the Utility's curves (section \ref{ConvertingFMWM}). As has already been mentioned, volumes we add to reconcile Utility's prices do not influence the shape of the demand curve in our Fundamental Model. Therefore, these volumes only affect the position of the red curve relative to the blue curve, but not the correlation between these two curves.}

 
\subsection{Determining the demand elasticity} 
Ultimately, an auxiliary study of demand elasticity was conducted.  
%
{Given that our optimization model is non-linear, our demand elasticities, too, should be non-linear. Estimating slope coefficients by means of a linear regression does not allow non-elastic components to be traced. Therefore, we will apply a finite central difference method to determine demand elasticities. To calculate the slope, we will rely on a function which depends on two parameters $p$ and $h$. The former parameter is a point on the demand curve, the latter one is a shift distance form the point $p$ and is equal to 100 MW. Then, {
	\begin{align}
	ls(p, h)=\frac{\FDem\left(\FDem^{-1}(p)+h\right)-\FDem\left(\FDem^{-1}(p)-h\right)}{2h},\label{LSCOEF}
	\end{align}
	where $ls(p,h)$ approximates the slope of the demand curve in point $p$ and is measured in EUR/MWh$^2$}. An elasticity coefficient can thus be determined as follows
		\begin{align}
	E_{F,i}^p=\frac{p}{\FDem^{-1}(p)}\cdot \frac{1}{ls(p, h)}. \label{lsS}
	\end{align}
	We have then conducted the study at points {$p \in \{0,20,25,30,35,40,50, 60\}$} and summarized the obtained results in Figures \ref{ELASTBarLsInelast} and \ref{ELASTBarInelast}. The former one shows average values of $ls(p,h)$, the latter one depicts the corresponding average elasticities $E_{F}^p$. 
	
	{Let us focus on both Figures simultaneously because the values of $ls(p,h)$ and $E_F^p$ follow similar patterns. First, note that $ls(p,h)$ at points $p \in \{0, 20, 25\}$ tend to be further away from zero. In turn, the absolute values of $E_F^p$ at points $p \in \{0, 20, 25\}$ tend to be relatively low. These two observations are consistent with the fact that the lower part of our fundamental demand curve is almost perfectly inelastic. 
	Second, absolute values of $ls(p,h)$ and $E_F^p$ at points $p \in \{30,35,40,45,50\}$ tend to be lower and higher, respectively, because these points are located in the most elastic parts of the fundamental demand curve. Moreover, the majority of equilibrium prices occurs within this price range. 
		
	More importantly, we see two spikes in the values of $ls(p,h)$ and $E_F^p$ at points $p \in \{30,35,40,45,50\}$. These spikes occur during morning and evening hours. Similar tendencies have also been observed by \cite{knaut2016hourly} and \cite{bigerna2014electricity}. However, contrary to these papers, absolute values of our elasticity coefficients drop. This tendency, too, is not surprising, given the composition of our fundamental demand curve. Points which split our fundamental demand curve into elastic and inelastic parts are typically located within the range $p \in \{30,50\}$. Therefore, following equation \ref{LSCOEF}, the difference $\FDem\left(\FDem^{-1}(p)+h\right)-\FDem\left(\FDem^{-1}(p)-h\right)$ may be large if point $p$ is located in the very elastic part of our fundamental demand. On the other hand, the values of $E^p_F$ are small if $p$ is located in its inelastic part. Hence, following section \ref{ModelDescription}, we may conclude that many demand elasticities were transferred to the supply side at the spike points. From this perspective, our results are similar to those of \cite{knaut2016hourly} and \cite{bigerna2014electricity}. 
	
	In fact, we can compare our results with only two studies on demand elasticity in the German wholesale electricity market. Our values {(range from -0.207 to -0.01)} are higher (in absolute terms) than those of \cite{knaut2016hourly} {(range from -0.006 to -0.001}), yet follow a similar pattern. On the other hand, our values are lower (in absolute terms) than the value $-0.43$ computed by \cite{bonte2015price}. Naturally, the discrepancy between the values stems from the method we used to determine demand elasticity. \cite{knaut2016hourly} and \cite{bonte2015price} used wind speed as the key parameter in their analyses. On the contrary, our values are non-linear and are obtained from an innovative fundamental model. Furthermore, the considered time frames, too, are different. }


{	To carry out a deeper sensitivity analysis, Figures \ref{ELMONTH} and \ref{ELWEEK} were constructed. The former one summarizes average monthly elasticities for selected prices $p$, the latter one demonstrates day-of-the-week elasticities. To the best of our knowledge, seasonal and day-of-the-week breakdown of demand elasticities has never been conducted for the German wholesale electricity market. 
	
As can be seen from Figure \ref{ELMONTH}, there is a spike in demand elasticity in summer. Following the reasoning in the previous paragraph, the observed tendency is consistent with  \cite{bigerna2014electricity} and does not seem counterintuitive. Finally, Figure \ref{ELWEEK} shows that elasticities tend to be higher during weekdays and lower over the weekend for higher values of $p$. The contrary can be said for lower values of $p$. The observed behavior can be well explained by routine electricity consumption patterns as described in e.g. \cite{weron2007modeling}. 
}

\section{Conclusion} \label{Conclusion} 
Auction curves recorded in a wholesale electricity market often contain a variety of arbitrage orders. The model developed in this paper attempts to detect and process these arbitrage orders. As a result, we can determine a true shape of the wholesale demand curve, i.e. the curve which is "cleaned" from those arbitrage orders. 
Our demand curve thus lies "in between" the elastic wholesale demand curve and a perfectly inelastic demand curve, the latter being obtained by transferring all elasticities to the supply side. 

Our model allowed us to obtain a more profound understanding of the German wholesale electricity market. In particular, we scrutinized the compositions of the wholesale demand and supply curves. The knowledge of these compositions may be useful for further modeling of the auction curves. Moreover, we showed that our model can approximate actual load values observed in an electricity market better than the model with perfectly inelastic demand curve. Therefore, the solution we elaborated can be well suitable for load forecasting. {Furthermore, our paper provides presumably the second (after \cite{knaut2016hourly}) in-depth analysis of demand elasticity in the German wholesale electricity market. }

{There exist numerous possibilities for further research. First, our model can be extended from the case of a market with single Utility, Retailer,  and Supplier. One can easily incorporate e.g. multiple Utilities, several types of Retailers, or differently acting Suppliers into the model. Second, one can smoothen the transition from elastic to inelastic parts in our fundamental demand curve. Currently there is a single point which splits this curve into two parts. Third, transaction costs can be considered. Their presence may substantially alter the behavior of market participants. Finally, the model can be extended with multiple coefficients, especially for determining the optimal amount of must-run supply and demand. As was mentioned in section \ref{ConvertingFMWM}, volumes of sell orders at price $\pmin$ and of buy orders at price $\pmax$ are much larger than those of other orders. Therefore, must-run orders can be treated differently. In fact, knowing the size of must-run buy and sell orders would allow us to omit the manipulations described in the second step of section \ref{ConvertingFMWM}. Shifting one of the Utility's curves to reconcile equilibrium prices will no longer be necessary because the optimal sizes of must-run supply and demand would be determined by the model.}

	\begin{figure}[h!]
	\hspace*{1cm}
	\parbox{2.2in}{%
		\hspace{-0.5cm}\subfigure[\normalsize Slope coefficients $ls(p, h)$ \label{ELASTBarLsInelast}]{\scalebox{0.9}{\input{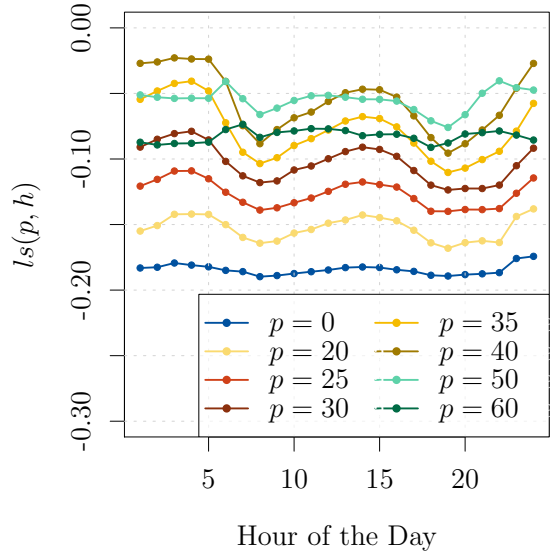}}}}
	\begin{minipage}{01in}%
		\hspace*{2.3cm}\subfigure[\normalsize Point elasticities of demand $E_{F}^p$ \label{ELASTBarInelast}]{\scalebox{0.9}{\input{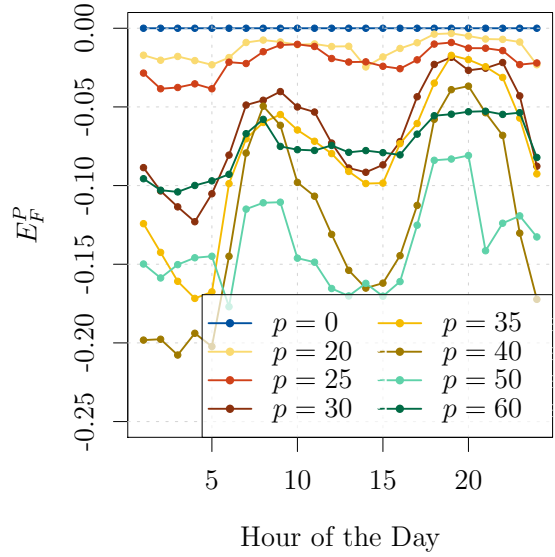}}}
	\end{minipage}
	
	\hspace*{1cm}
	\parbox{2.2in}{%
		\hspace{-0.5cm}\subfigure[\normalsize Average monthly elasticities \label{ELMONTH}]{\scalebox{0.9}{\input{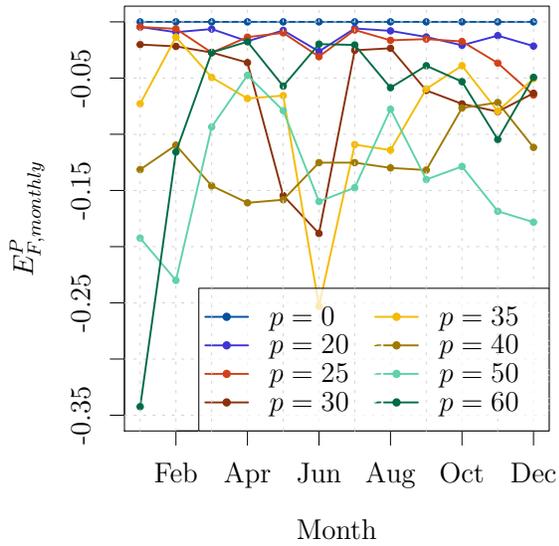}}}}
	\begin{minipage}{01in}%
		\hspace*{2.3cm}\subfigure[\normalsize Day-of-the-week elasticities \label{ELWEEK}]{\scalebox{0.9}{\input{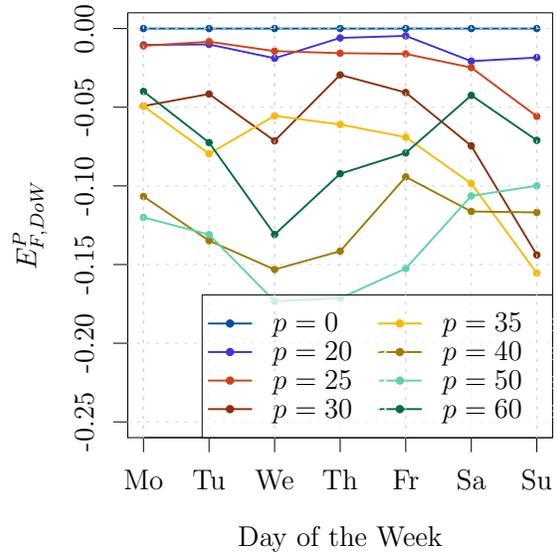}}}
	\end{minipage}
	
	\vspace{0.2cm}
	\caption{The analysis of demand elasticities}
	\label{ElMain}
\end{figure}

\newpage

\newpage 
\bibliographystyle{apalike}
\bibliography{lib}

\end{document}